\documentclass[11pt, aps, prd, reprint, onecolumn, sort&compress, superscriptaddress, showpacs, nofootinbib, preprintnumbers]{revtex4-1}
\pdfoutput=1
\usepackage{graphicx}
\usepackage{dcolumn}
\usepackage{bm}
\usepackage{amsmath}
\usepackage{amssymb}
\usepackage{dsfont}
\usepackage{amsfonts}
\usepackage[ngerman,UKenglish]{babel}
\usepackage{hyperref}
\usepackage{nicefrac}
\usepackage{verbatim}
\usepackage{bbm}
\usepackage{color}

\def\dbar{{\mathchar'26\mkern-11mu  d}}

\newcommand{\bx}{\boldsymbol{x}}
\newcommand{\by}{\boldsymbol{y}}

\newcommand{\bk}{\boldsymbol{k}}
\newcommand{\bp}{\boldsymbol{p}}
\newcommand{\bq}{\boldsymbol{q}}

\newcommand{\bg}{\boldsymbol{\gamma}}
\renewcommand{\vec}[1]{\mbox{\boldmath$#1$\unboldmath}}

\begin{document}

\title{Effect of Transverse Gluons on Chiral Restoration in Excited Mesons}
\date{\today}
\author{M.~Pak}
\email{markus.pak@uni-graz.at}
\author{L.~Ya.~Glozman}
\email{leonid.glozman@uni-graz.at}
\affiliation{Institut f\"ur Physik, FB Theoretische Physik, Universit\"at Graz, Universit\"atsplatz 5,
8010 Graz, Austria}

\begin{abstract}
The effect of transverse gluons on the chiral symmetry patterns of
excited mesons is studied in a Coulomb gauge QCD model. The linear rising static quark-antiquark potential and the transverse gluon propagator known from lattice
studies are input into the model. The non-perturbative quark propagator, which enters the 
meson bound state equations, is derived
from the Dyson--Schwinger equations and a complete set of mesons for general spin quantum number is presented. 
From analyzing the bound state equations for large spins it is demonstrated,
that chiral and axial symmetry are restored. In this limit a complete degeneracy of all multiplets with given spin 
is observed. The effect of the transverse gluon interaction
is shown to vanish rapidly as the spin quantum number is increased. For vanishing dynamical quark mass
the expected meson degeneracies are recovered. 
\end{abstract}


\maketitle

\section{Introduction}
Understanding the two cornerstones of non-perturbative QCD - confinement and chiral symmetry breaking -
is one of the most important challenges of modern particle physics. The hadron spectrum offers a unique 
possibility to get insight into the underlying mechanisms of both these phenomena and their interplay.
For instance, we certainly know that for the low energy quark sector of QCD the chiral symmetry and its spontaneous breaking are
crucially important.
Quark condensation in the vacuum breaks the $U_{\textsc{L}}(N_{\textsc{F}}) \times U_{\textsc{R}}(N_{\textsc{F}})$
symmetry group of massless QCD to the vector subgroup $U_{\textsc{V}}(N_{\textsc{F}})$, 
preventing the ground state parity partners to have equal masses and giving rise to $N^2_{\textsc{F}}$
pseudoscalar Goldstone bosons. The mass of the $\eta'$ particle is shifted up due to the anomalous breaking of the $U_{\textsc{A}}(1)$ symmetry.  

However, higher states in the light hadron spectrum show a different picture: mesons and baryons tend to form
chiral multiplets, signaling the ``effective'' restoration 
of the $U_{\textsc{A}}(N_{\textsc{F}})$ symmetry \cite{Glozman:1999tk,*Cohen:2001gb,*Glozman:2002cp,*Glozman:2003bt}, 
see Ref.~\cite{Glozman:2007ek} for an overview. 
However, for a final proof missing states have
to be experimentally confirmed. The physical picture is the following: for high-lying states the valence
quarks should become unaffected by the condensate, since one enters
the semi-classical regime, where the quantum fluctuations are suppressed, see Ref.~\cite{Glozman:2004gk, *Glozman:2005tq}. 
This phenomenological picture has been tested, however, with only a
static quark-antiquark Lorentz-vector confining potential entering the system, which has either been chosen to be
of harmonic oscillator-type \cite{Kalashnikova:2005tr} or linearly rising \cite{Wagenbrunn:2006cs}. 
Here we close a gap and show that highly excited high-spin mesons form approximately degenerate chiral 
multiplets also for a transverse gluon type of interaction.

The Nambu--Jona-Lasinio type mechanism of spontaneous breaking of chiral symmetry has been successfully analyzed in effective potential models, 
Refs.~\cite{Finger:1981gm, LeYaouanc:1983iy, Adler:1984ri}, however, with the low energy chiral properties
of the theory predicted too low. It has been observed that an additional transverse
gluon exchange increases the values of the quark condensate and dynamical quark mass substantially towards
their phenomenological values, Refs.~\cite{Alkofer:1988tc, Szczepaniak:2002ir, LlanesEstrada:2004wr, Pak:2011wu, Fontoura:2012mz}. Most recently, in Ref.~\cite{Pak:2011wu} this
has been shown 
by applying the variational approach to the quark sector of Coulomb gauge QCD. 
Since transverse gluons have a significant effect on chiral symmetry breaking we here ask the question if they also 
affect high-spin mesons, where chiral symmetry is expected to be effectively restored.

In Refs.~\cite{Wagenbrunn:2006cs, Wagenbrunn:2007ie} it has been shown that in a model
with chiral quarks and a color-Coulomb linear potential mesons fall into parity
multiplets for large spins. Here we generalize the model and include transverse gluons. We demonstrate that the
additional interaction respects chiral $U_{\textsc{L}}(2) \times U_{\textsc{R}}(2)$ symmetry as well. 

The organization of the paper is as follows: in Section
\ref{Hamiltonian-sec} we review the model Hamiltonian
generalized to transverse gluon interaction and specify the interaction
kernels from lattice Coulomb gauge studies. In Section
\ref{Quark-Prop-Chapter} the quark propagator dressing functions are
obtained from the Dyson--Schwinger equations in a rainbow truncation and 
the infrared divergencies of these quantities are discussed. The symmetry properties
of non-interacting quarks are listed in Section \ref{Multipl-chap}.
A complete spectrum of meson
Bethe--Salpeter equations for arbitrary orbital quantum number are presented in Section \ref{Bethe}. The
limit of vanishing dynamical quark mass is discussed in Section \ref{Mass-zero-chap}. In Section
\ref{Restor} the equations are studied for large spins $J$. It is shown by numerically analyzing the
angular integrals of the coupled system that the effect of
transverse gluons rapidly vanishes for increasing spin. Effective chiral
symmetry restoration is found to persist for such an interaction. In Section \ref{XIIsum} we
summarize our main findings. Irreducible tensor relations
and the Bethe--Salpeter equations are collected in Appendices 
\ref{tensor-rel-chap} and 
\ref{Appendix-coupled}.

\section{Model Hamiltonian}
\label{Hamiltonian-sec} 
\begin{figure}[t]
\centering
\includegraphics[width=.4\linewidth]{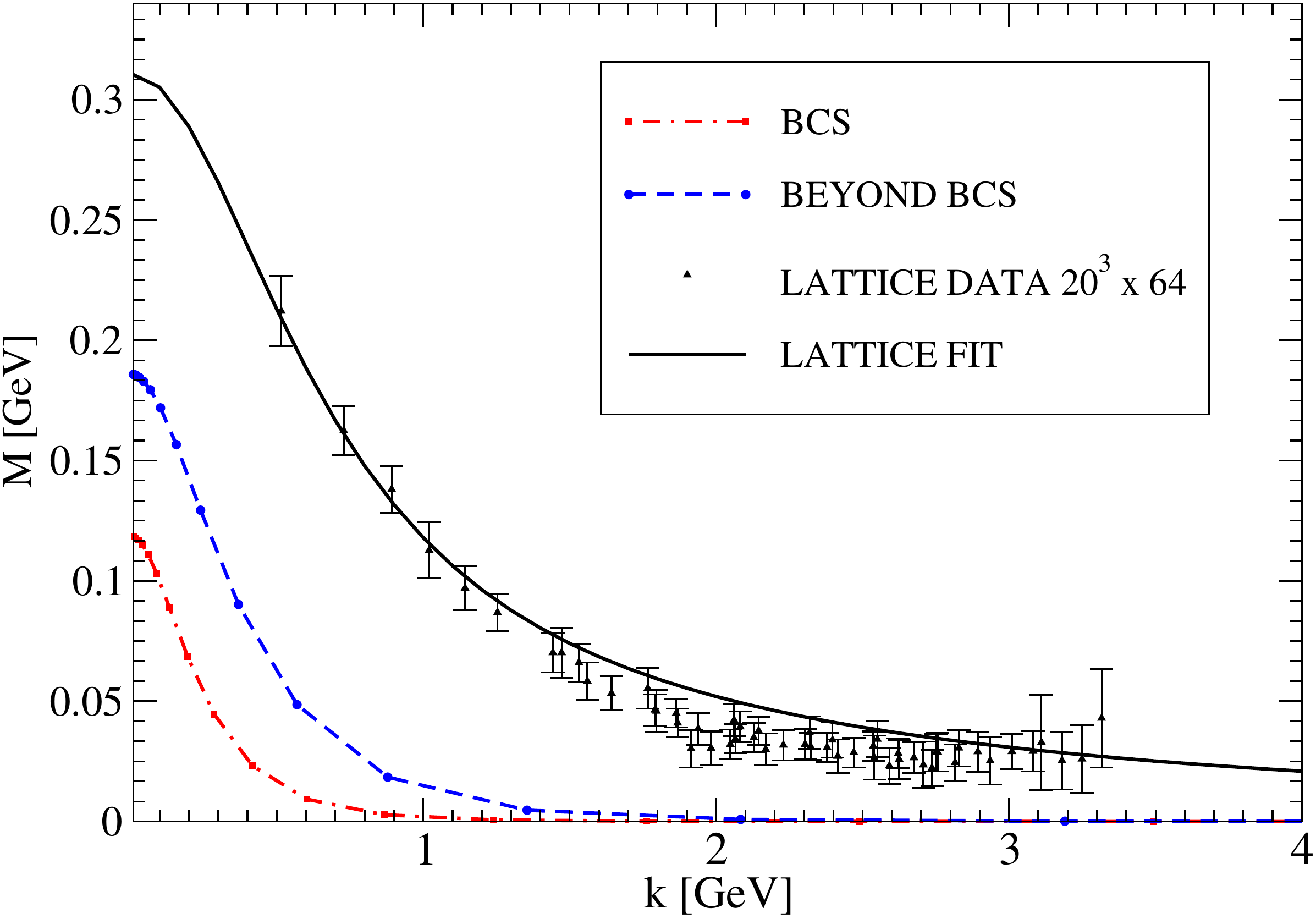}
\includegraphics[width=.4\linewidth]{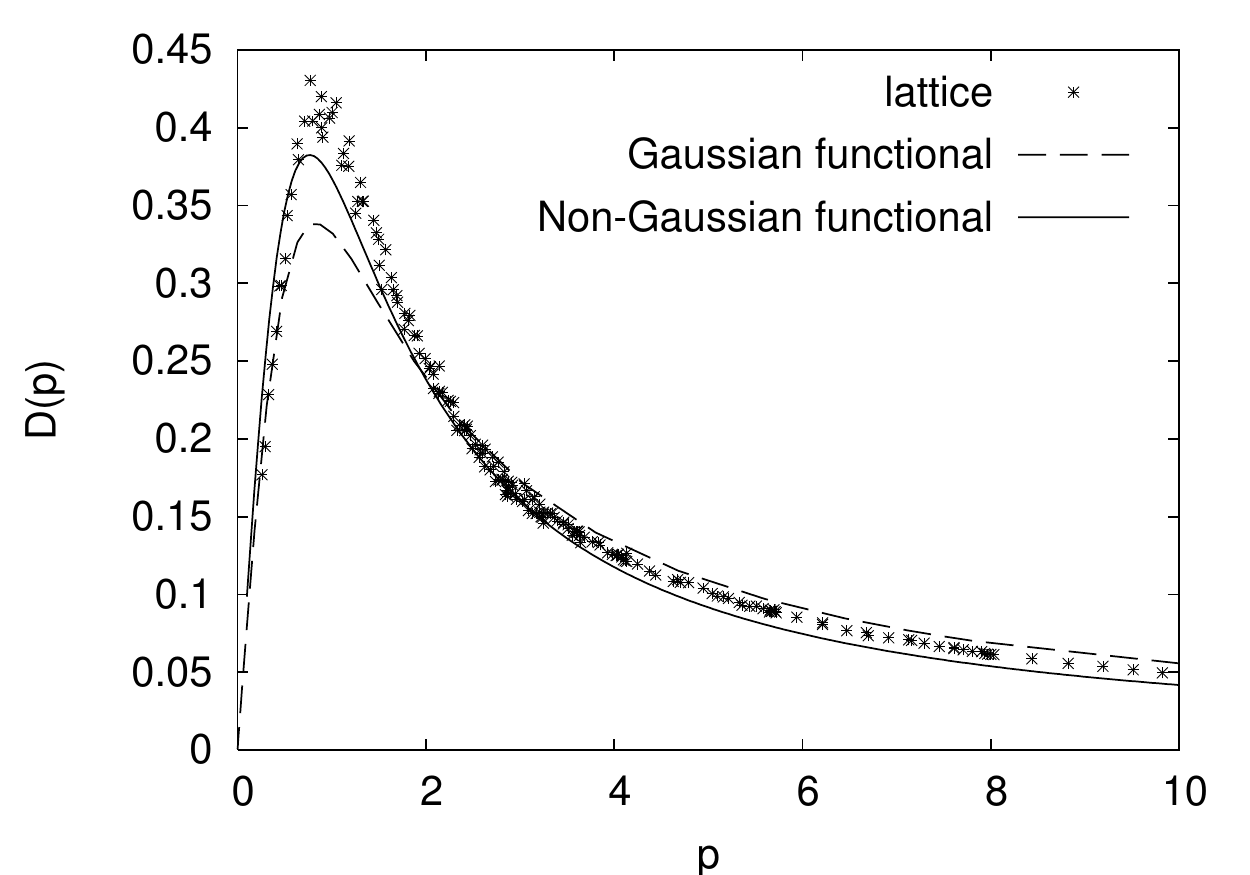}
\caption[Dynamical mass compared to lattice]{\sl
\textit{Left Panel:} Dynamical quark mass for $\sigma_{\textsc{C}} = 2 \sigma_{\textsc{W}}$ and $\sqrt{\sigma_{\textsc{W}}} = 440 $ MeV
comparing the results from the variational approach to QCD in Coulomb gauge, Ref.~\cite{Pak:2011wu}, to the lattice data obtained in Ref.~\cite{Burgio:2012ph}.  
The dashed curve is obtained with the additional coupling of quarks to transverse gluons while the dash-dotted curve is obtained using the color-Coulomb potential only, 
for details see Ref.~\cite{Pak:2011wu}. \textit{Right Panel:} Instantaneous transverse gluon propagator $D(p)$. Data points show the lattice results
of Ref.~\cite{Burgio:2008jr}. The dashed curve is the result of the variational analysis using a Gaussian vacuum, Ref.~\cite{Feuchter:2004mk} 
and the full curve shows the extension to
non-Gaussian wave functionals. The plot is taken from Ref.~\cite{Campagnari:2010wc}.
} \label{fig-dynam-mass-lattice}
\end{figure}

The model Hamiltonian generalized to
transverse gluons is defined as (see Refs.~\cite{Wagenbrunn:2007ie, Alkofer:1988tc})
\begin{align}
 H = H_{\textsc{F}} + H_{\textsc{I}} \; , \qquad H_{\textsc{I}} = H_{\textsc{C}} +
 H_{\textsc{T}} \; ,
\label{model-Ham}
\end{align}
where
\begin{align}
\label{167-x4}
H_{\textsc{F}} = \int d^3 x \, \psi^\dagger (\bx) \left( -i  \vec{\alpha} \cdot \boldsymbol{\partial} + \beta m_0 \right) \psi (\bx)
\end{align}
is the free Hamiltonian of the quark field $\psi (\bx)$.
Here $\vec{\alpha}, \beta$ are the usual Dirac matrices satisfying
$\{\alpha_i,\alpha_j\} = \delta_{i j}$ and $\{\beta, \alpha_i \} =
0$, $m_0$ is the current quark mass. The interaction Hamiltonian
$H_{\textsc{I}}$ is split up into the instantaneous linear confining Coulomb term
\begin{align}
\label{Ham-Coul} H_{\textsc{C}} = \frac{1}{2} \int d^3 x \, d^3 y
\, \psi^{\dagger} (\bx) T^a \psi(\bx) \, V_{\textsc{C}}^{ab} (|\bx-\by|) \,
\psi^{\dagger} (\by) T^b \psi(\by) \; ,
\end{align}
with $T^a$ being the Hermitian generators of the gauge group $SU(N)$ in the fundamental representation 
and the transverse gluon interaction of the form
\begin{align}
\label{Ham-trans}
 H_{\textsc{T}} = - \frac{1}{2} \int d^3 x \,  d^3 y \, \psi^{\dagger} (\bx)\alpha_i T^a \psi(\bx) \,
  D_{i j}^{ab} (|\bx-\by|) \, \psi^{\dagger} (\by) \alpha_j T^b \psi(\by) \; .
\end{align}
Here a comment is in order: the interaction Hamiltonian (\ref{Ham-Coul}) arises naturally in Coulomb gauge QCD from the kinetic energy of the longitudinal
modes after resolving Gauss' law. 
Using a perturbative gluon propagator and integrating out the gluonic degrees of freedom leads to an additional quark 
interaction due to transverse gluons, Eq.~(\ref{Ham-trans}). A recent Coulomb gauge study supports such an interaction kernel. 
Using the so-called variational approach to QCD in Coulomb gauge, a quark 
vacuum wave functional is suggested, which goes beyond the typical BCS approximation and includes a coupling
between quarks and transverse gluons, see Ref.~\cite{Pak:2011wu}. The additional transverse gluon interaction increases
the chiral condensate and the constituent quark mass of about $20 -
60 \%$. These findings are in agreement with the study of
Ref.~\cite{Fontoura:2012mz}, where a potential of the form
(\ref{Ham-trans}) is used to describe transverse gluons. For a gluon propagator 
which vanishes for small momenta, both these approaches lead to a constituent quark mass of $\approx 200$ MeV. In Fig.~\ref{fig-dynam-mass-lattice} 
on the left-hand side the dynamical quark mass is plotted and compared
to the cases without the additional transverse gluon interaction and to most recent lattice Coulomb gauge data, see Ref.~\cite{Burgio:2012ph}.  
For the purpose of our work it is much more convenient to use a model
Hamiltonian as considered in Eq.~(\ref{Ham-trans}). 

We now specify the interaction kernels $V^{ab}_{\textsc{C}}(|\bx-\by|)$ and $D_{i j}^{ab}(|\bx-\by|)$, appearing in Eqs.~(\ref{Ham-Coul}), (\ref{Ham-trans}). Both are diagonal
in color space. From the variational approach to Coulomb gauge, Refs.~\cite{Feuchter:2004mk, Epple:2006hv}, one finds a potential which at large distances rises linearly
\begin{align}
\label{389}
V_{\textsc{C}} (r) = \sigma_{\textsc{C}} r \, , \, r \to \infty \; , \qquad r = |\bx-\by| \, .
\end{align}
The same behavior is found on the lattice, Ref.~\cite{Voigt:2008rr,*Nakagawa:2011ar,*Greensite:2003xf}, with a Coulomb string tension $\sigma_{\textsc{C}}$ of
\begin{align}
\label{394}
\sigma_{\textsc{C}} \approx \left( 2 \ldots 3 \right) \sigma_{\textsc{W}} \, ,
\end{align}
where $\sqrt{\sigma_{\textsc{W}}} = 440$ MeV is the Wilsonian string
tension. 
It is well known that the linearly rising confinement potential
$V_{\textsc{C}}(r)$ in Eq.~(\ref{Ham-Coul}) leads to infrared
divergences in momentum space, which, however, cancel from the observable
quantities as shown in Ref.~\cite{Adler:1984ri}. For the actual
computation we introduce an (infrared) cut-off
parameter, which, in the end of the calculation, is send to zero.
For the definition of the infrared regular confinement potential
we adopt the convention of Refs.~\cite{Alkofer:1988tc,
Wagenbrunn:2007ie}, given as (in momentum space)
\begin{align}
 V_{\textsc{C}}(\bp) = \frac{8 \pi \sigma_{\textsc{C}}}{(\bp^2+\mu^2_{\textsc{IR}})^2} \; ,
\end{align}
with $\sigma_{\textsc{C}}$ the Coulomb string tension
and $\mu_{\textsc{IR}}$ the infrared cut-off parameter. Due to asymptotic freedom the color-Coulomb potential also has
an ordinary short-range $1/r$ part, which influences the UV-part of the dynamical quark mass. Since 
we want to investigate the role of transverse gluons for chiral symmetry restoration in highly excited mesons, the Coulomb
potential is not required and can be neglected.

As static transverse gluon propagator $D(\bp)$ in
Eq.~(\ref{Ham-trans}) we use the result from lattice studies, Ref.~\cite{Burgio:2008jr} in
Coulomb gauge QCD, which can nicely be fitted by Gribov's formula
\begin{align}
\label{transv-gluon}
 D_{i j}^{ab} (\bp) = \delta^{a b} t_{ij}(\bp) D(\bp) \; , \qquad \text{with} \qquad D^{-1}(\bp) = 2 \, \sqrt{\bp^2+\frac{M_{\textsc{G}}^4}{\bp^2}} \; ,
\end{align}
where $M_{\textsc{G}} \approx 880 $ MeV is a mass scale referred to as Gribov mass and the transverse
projector is defined as 
\begin{align} t_{ij}(\bx) = \int \dbar^3 p \, \left(\delta_{ij} - \hat{p}_i \hat{p}_j \right) e^{i \bp \cdot \bx} \; , \quad \dbar^3 q \equiv \frac{d^3
q}{(2\pi)^3} \; , \quad \hat{\bp} = \frac{\bp}{|\bp|} \; . 
\end{align}
On the right-hand side of Fig.~\ref{fig-dynam-mass-lattice} the transverse gluon propagator is presented and compared to results in the variational approach to QCD, 
Refs.~\cite{Feuchter:2004mk, Campagnari:2010wc}. 

\section{Quark Propagator and Gap Equation}
\label{Quark-Prop-Chapter} Due to the static interaction kernels (\ref{Ham-Coul}), (\ref{Ham-trans}), the dressed inverse quark propagator can be
decomposed as
\begin{align}
\label{quark-dressing}
 i S^{-1}(p_0, \bp) = \gamma_0 p_0 - \bg \cdot \hat{\bp} B(\bp) - A(\bp) \;
 ,
\end{align}
with $A(\bp), B(\bp)$ denoting scalar and vector dressing
functions. Within this large-$N_{\textsc{C}}$ model only the rainbow diagrams contribute to 
the gap equation (see
Refs.~\cite{LeYaouanc:1983iy, Adler:1984ri} for a detailed derivation)
\begin{align}
\label{A}
 A(\bp) &= m + \frac{1}{2} \int \dbar^3 q \, \left[ V(\bk) + 2 D(\bk) \right] \frac{A(\bq)}{\omega(\bq)} \; , \quad \\
\label{B} 
B(\bp) &= |\bp| + \frac{1}{2} \int \dbar^3 q \, \left[ V(\bk)
(\hat{\bp} \cdot \hat{\bq}) + 2 D(\bk) (\hat{\bp} \cdot \hat{\bk})
(\hat{\bq} \cdot \hat{\bk}) \right] \frac{|\bq|}{\omega(\bq)} \; , 
\end{align}
with $\bk = \bp - \bq$ and $\omega^2(\bp) = A^2(\bp) + B^2(\bp)$. We stress that 
the quark-gluon vertex is taken to be bare and it is assumed that no infrared divergence occurs in the transverse gluon sector. Spontaneous
chiral symmetry breaking reflects itself in a non-vanishing scalar dressing function $A(\bp)$ and in an infrared finite dynamical quark mass
function $M(\bp)$, which is defined as (see Fig.~\ref{fig-dynam-mass-lattice})
\begin{align}
 M(\bp) = |\bp| \frac{A(\bp)}{B(\bp)} \; .
\end{align}
The plateau value at zero momentum $M(0)$ is interpreted as constituent quark mass. For highly excited states, where the typical quark momentum is large,
the valence quarks become less affected by the
quark condensate, hence  
in this regime the behavior of the quark dressing functions (\ref{A}), (\ref{B}) for large momenta $p$ is essential. 
At tree-level in the chiral limit they
approach
\begin{align}
\label{tree-level}
\qquad A(\bp \rightarrow \infty) \rightarrow 0 \; , B(\bp \rightarrow \infty) \rightarrow |\bp| \; , 
\end{align}
and the dynamical quark mass $M(\bp)$ goes to zero. 
For later purpose it is important to project out the infrared
divergences, which has explicitly been done in
Ref.~\cite{Wagenbrunn:2007ie} for the case of a purely linear color-Coulomb potential and is not changed by the additional
transverse gluon interaction. We therefore quote the result of Ref.~\cite{Wagenbrunn:2007ie}, given as
\begin{align}
\label{A-infrared}
 A(\bp) &= \frac{\sigma}{2\mu_{\textsc{IR}}} \frac{A(\bp)}{\omega(\bp)} + A_{\textsc{f}}(\bp) \; , \\
\label{B-infrared}
 B(\bp) &= \frac{\sigma}{2\mu_{\textsc{IR}}} \frac{|\bp|}{\omega(\bp)}+ B_{\textsc{f}}(\bp) \; , \\
\label{omega-infrared}
 \omega(\bp) &= \frac{\sigma}{2\mu_{\textsc{IR}}} + \omega_{\textsc{f}}(\bp) \; , 
\end{align}
with $A_{\textsc{f}}(\bp), B_{\textsc{f}}(\bp),
\omega_{\textsc{f}}(\bp)$ being infrared-finite.

\section{Symmetry Properties of Non-Interacting Quarks}
\label{Multipl-chap}
Before we proceed with deriving the meson bound state equations, we discuss the expected degeneracies in case of
non-broken chiral symmetry. 
The two-quark amplitudes then follow from Poincar\'{e} invariance and $U_{\textsc{L}}(2) \times U_{\textsc{R}}(2)$ symmetry. 
The complete list of $\overline{q} q$-meson $SU_{\textsc{L}}(2) \times SU_{\textsc{R}}(2)$ multiplets is given as 
(see Refs.~\cite{Glozman:2002cp, Glozman:2003bt, Glozman:2007ek} for 
a detailed explanation) ($k=1,2,\ldots$)
\begin{align}
\label{mes-deg}
\boldsymbol{J = 0:} \qquad \qquad \; \; 
&\begin{cases}(1/2,1/2)_a : &\quad1, 0^{-+} \longleftrightarrow 0, 0^{++}\\
(1/2,1/2)_b : &\quad1, 0^{++} \longleftrightarrow 0, 0^{-+}
\end{cases}\\
 \boldsymbol{J=2k:} \qquad \qquad
&\begin{cases} 
(0,0) : &0, J^{--} \longleftrightarrow 0, J^{++}\\
(1/2,1/2)_a : &1, J^{-+} \longleftrightarrow 0, J^{++}\\
(1/2,1/2)_b : &1, J^{++} \longleftrightarrow 0, J^{-+}\\
(0,1) \oplus (1,0) : &1, J^{++} \longleftrightarrow 1, J^{--}
\end{cases}\\
\boldsymbol{J=2k-1:} \qquad
&\begin{cases} 
(0,0) : &0, J^{++} \longleftrightarrow 0, J^{--}\\
(1/2,1/2)_a : &1, J^{+-} \longleftrightarrow 0, J^{--}\\
(1/2,1/2)_b : &1, J^{--} \longleftrightarrow 0, J^{+-}\\
(0,1) \oplus (1,0) : &1, J^{--} \longleftrightarrow 1, J^{++} \; .
\label{mes-deg3}
\end{cases}
\end{align}
The axial $U_{\textsc{A}}(1)$ symmetry mixes the states from the $(1/2,1/2)_a$ and $(1/2,1/2)_b$ chiral multiplets with the same isospin but opposite parity. 
Moreover, from theoretical arguments, Ref.~\cite{Glozman:2002kq}, and from the analysis with a color-Coulomb potential in the limit $J\rightarrow \infty$, Ref.~\cite{Wagenbrunn:2007ie}, 
one expects an even higher degree of degeneracy
\begin{align}
 \label{chir-mul}
[(0,1/2) \oplus (1/2,0)] \times [(0,1/2) \oplus (1/2,0)] \; , 
\end{align}
i.e., all states with given spin should become degenerate. 

Next we turn to the interacting case, where the quark self-energy breaks chiral and $U_{\textsc{A}}(1)$ axial symmetry. 
Chiral symmetry restoration then means that the splittings within the multiplets vanish in the limit $J\rightarrow \infty$. 
It will be interesting to observe if the 
transverse gluon interaction still allows for such a degeneracy for large orbital quantum numbers. 
To investigate the effect of transverse gluons we derive bound state equations for the model Hamiltonian (\ref{model-Ham}). 

\section{Bethe--Salpeter Equations for Mesons}
\label{Bethe} The homogeneous ($\overline{q} q $)-meson Bethe--Salpeter equation (BSE) with total and relative four momenta $P, p$
and individual momenta $q\pm P/2$ of the (anti-) quarks is
conveniently set up in momentum space
\begin{align}
 \chi(P,p) = -i \int \frac{d^4 q}{(2\pi)^4} \, K(P,p,q) \, S(q+P/2) \chi(P,q) \,
S(q-P/2) \: ,
\end{align}
with $K(P,p,q)$ the Bethe--Salpeter kernel, $S(q\pm P/2)$ the
quark propagator and $\chi(P,p)$ the meson vertex function. Dirac and color indices are suppressed. 
As a first step one usually chooses the rest frame, where the (anti-) quark momenta have
equal magnitude, so that $P^{\mu} = (\mu, \boldsymbol{P} = 0)$. Due to the
instantaneous interactions of our model the Bethe--Salpeter kernel and the vertex function depend only on
three momentum and the BSE (in the ladder approximation, which is exact within this large-$N_{\textsc{C}}$ model) simplifies as (see
Ref.~\cite{Alkofer:1988tc})
\begin{align}
\label{BSE}
 \chi(\mu,\bp) = -&i \int \frac{d^4 q}{(2\pi)^4} \, V(\bk) \gamma_{0} \,
 S(q_0+\frac{\mu}{2},\bq) \chi(\mu,\bq) \,
S(q_0-\frac{\mu}{2},\bq) \, \gamma_{0}  \nonumber \\
+ & \, i  \int \frac{d^4 q}{(2\pi)^4} \,D_{ij}(\bk) \gamma_{i} \,
 S(q_0+\frac{\mu}{2},\bq) \chi(\mu,\bq) \,
S(q_0-\frac{\mu}{2},\bq) \, \gamma_{j} \: .
\end{align}
This equation is solved in the standard fashion: the $q^0$-integral is taken explicitly, the vertex function $\chi(\mu,\bq)$ of a given meson type expanded
in terms of all possible Poincar\'{e}-invariant amplitudes (see Appendix \ref{Appendix-coupled}) and, by applying appropriate traces of Dirac matrices, 
 the independent tensor components $\chi_i$ (see Eqs.~(\ref{Cat1})-(\ref{Cat3})) are projected out. Since we are interested in meson bound states for general orbital excitations $J$, 
irreducible tensor products occur in the construction of the independent vertex components, 
which can be simplified by using formulas listed in Appendix \ref{tensor-rel-chap}.

As in Ref.~\cite{Wagenbrunn:2007ie} three different types of mesons with total angular
momentum, parity and $C$-parity $J^{PC}$ are constructed
\begin{align}\nonumber &\mbox{Category 1:} \qquad \begin{cases}
        J^{-+}, & J=2n\\
        J^{+-}, & J=2n+1
       \end{cases}\\ \nonumber
&\mbox{Category 2:} \qquad \begin{cases}
 J^{++}, & J=2n\\
        J^{--}, & J=2n+1
              \end{cases}\\ \nonumber
&\mbox{Category 3:} \qquad \begin{cases}
        J^{--}, & J=2(n+1)\\
        J^{++}, & J=2n+1 \; .
       \end{cases}
\end{align}
A fourth possible category with $(2n)^{+-}, (2n+1)^{-+}$ is absent in a system with only instantaneous interactions. 
In Appendix \ref{Appendix-coupled} the coupled integral equations for all meson categories are presented. 

In Ref.~\cite{Wagenbrunn:2007ie} it has
been shown that in the limit $J\rightarrow \infty$ all states for a given $J$ are completely degenerate. Here we repeat the calculation for an additional transverse gluon interaction
and investigate if the same degree of degeneracy is still present.   
As a first step towards this study we have to analyze the BSEs regarding the infrared limit $\mu_{\textsc{IR}} \rightarrow 0$.
Mesons of category one are
described by two coupled integral equations, see Eqs.~(\ref{rev-1}), (\ref{rev-2}) ($\bk=\bp-\bq$)
\begin{subequations}
\begin{align}
\label{Mes1-1}
 \omega(\bp) h(\bp) &= \frac{1}{2} \int \dbar^3 q \, \left( V_{\textsc{C}}(\bk)+2 \, D(\bk) \right) P_{J}(\hat{\bp} \cdot \hat{\bq})
\left[ h(\bq) + \frac{\mu^2}{4 \omega(\bq)} g(\bq) \right] \; , \\
\label{Mes1-2}
\left[ \omega(\bp) - \frac{\mu^2}{4 \omega(\bp)} \right] g(\bp) &= h(\bp)  \nonumber \\
+ \frac{1}{2} \int \dbar^3 q V_{\textsc{C}}(\bk) &\left\{
\frac{A(\bp) A(\bq) P_{J}(\hat{\bp} \cdot \hat{\bq}) + B(\bp)
B(\bq)
\left(\frac{J+1}{2J+1} P_{J+1}(\hat{\bp} \cdot \hat{\bq})  + \frac{J}{2J+1} P_{J-1}(\hat{\bp} \cdot \hat{\bq}) \right)}{\omega(\bp) \omega(\bq)} \right\} g(\bq)  \nonumber \\
- \frac{1}{2} \int \dbar^3 q \, 2 \, D(\bk) & \left\{
\frac{A(\bp)A(\bq) P_{J}(\hat{\bp} \cdot \hat{\bq}) + B(\bp)B(\bq)
\left(\frac{J+1}{2J+1} F_1  + \frac{J}{2J+1}
G_1 \right)}{\omega(\bp) \omega(\bq)} \right\} g(\bq) \; , 
\end{align}
\end{subequations}
where we have defined the following quantities
\begin{align}
\label{Def-F1}
F_1 &=  \frac{(\hat{\bk} \cdot \hat{\bq})}{|\bk|} \left(
 |\bp|
P_{J}(\hat{\bp} \cdot \hat{\bq}) - |\bq| P_{J+1}(\hat{\bp} \cdot \hat{\bq}) \right) \; , \\
\label{Def-G1}
G_1 &=   \frac{(\hat{\bk} \cdot
\hat{\bq})}{|\bk|} \left( |\bp| P_{J}(\hat{\bp} \cdot \hat{\bq}) -
|\bq| P_{J-1}(\hat{\bp} \cdot \hat{\bq}) \right) \; ,
\end{align}
which appear for the transverse gluon interaction. With help of
Eqs.~(\ref{A-infrared}), (\ref{B-infrared}),
(\ref{omega-infrared}) we rewrite Eqs.~(\ref{Mes1-1}), (\ref{Mes1-2}) as
\begin{subequations}
\begin{align}
\label{Mes1-inf1}
 \omega_{\textsc{F}}(\bp) h(\bp) &= \frac{\mu^2}{4} g(\bp) + \frac{1}{2} \int \dbar^3 q \left(V_{\textsc{C},\textsc{F}}(\bk) +2 D(\bk)\right)  P_{J}(\hat{\bp} \cdot \hat{\bq}) h(\bq) \; , \\
\label{Mes1-inf2}
 \omega_{\textsc{F}}(\bp) g(\bp) &= h(\bp)  \nonumber \\
+ \frac{1}{2} \int \dbar^3 q V_{\textsc{C},\textsc{F}}(\bk)
&\left\{ \frac{A(\bp) A(\bq) P_{J}(\hat{\bp} \cdot \hat{\bq}) +
B(\bp) B(\bq)
\left(\frac{J+1}{2J+1} P_{J+1}(\hat{\bp} \cdot \hat{\bq})  + \frac{J}{2J+1} P_{J-1}(\hat{\bp} \cdot \hat{\bq}) \right)}{\omega(\bp) \omega(\bq)} \right\} g(\bq)  \nonumber \\
- \frac{1}{2} \int \dbar^3 q \, 2 \, D(\bk) & \left\{
\frac{A(\bp)A(\bq) P_{J}(\hat{\bp} \cdot \hat{\bq}) + B(\bp)B(\bq)
\left(\frac{J+1}{2J+1} F_1  + \frac{J}{2J+1}
G_1 \right)}{\omega(\bp) \omega(\bq)} \right\} g(\bq) \; .
\end{align}
\end{subequations}
In addition we have used that
\begin{align}
 \lim_{\mu_{\textsc{IR}}\rightarrow 0} \frac{\mu_{\textsc{IR}}}{\pi^2} \int d^3 q \, \frac{1}{(\bk^2+\mu^2_{\textsc{IR}})^2} f(\bq) = \int d^3 q \, \delta(\bp-\bq) f(\bq) = f(\bp) \: ,
\end{align}
$V_{\textsc{C},\textsc{F}}(\bk) = V_{\textsc{C}}(\bk) - \frac{\sigma_{\textsc{C}}}{\mu_{\textsc{IR}}} (2\pi)^3 \delta(\bk)$
and that the integral
\begin{align}
 \frac{1}{2} \int \dbar^3 q \, 2 D(\bk) P_{J}(\hat{\bp} \cdot \hat{\bq}) \frac{\mu^2}{4 \omega(\bq)} g(\bq) \sim \mathcal{O}(\mu_{\textsc{IR}}) \; ,
\end{align}
vanishes in the limit $\mu_{\textsc{IR}} \rightarrow 0$. We note, that the transverse gluon propagator $D(\bk)$, Eq.~(\ref{transv-gluon}), is finite for
$\mu_{\textsc{IR}} \rightarrow 0$. 

Four linearly independent equations
describe mesons in category two (\ref{Mes2-1})-(\ref{Mes2-4}) given in the infrared limit as ($P_J = P_{J}(\hat{\bp} \cdot \hat{\bq})$)
\begin{subequations}
\begin{align}
\omega_{\textsc{f}}(\bp) h_1(\bp) &=
\frac{\mu^2}{4} g_1(\bp)  \nonumber  \\ 
+ \frac{1}{2} \int \dbar^3 q \, & V_{\textsc{C},\textsc{f}}(\bk)
\Bigg\{ \left( \frac{J}{2J+1} P_{J+1}
+ \frac{J+1}{2J+1} P_{J-1}\right) h_1(\bq)  + \frac{A(\bq)}{\omega(\bq)}
\frac{\sqrt{J(J+1)}}{2J+1} \left(P_{J+1} -
P_{J-1}\right) h_2(\bq) \Bigg\}  \nonumber \\
- \frac{1}{2}  \int \dbar^3 q \, & 2 D(\bk) \Bigg\{ \left(
\frac{J}{2J+1} F_2
+ \frac{J+1}{2J+1} G_2 \right)  h_1(\bq) + \frac{A(\bq)}{\omega(\bq)}
\frac{\sqrt{J(J+1)}}{2J+1} \left( F_1 -
G_1 \right) h_2(\bq) \Bigg\} \; , \label{Mes2-inf1} \\
 \omega_{\textsc{f}}(\bp)
 g_1(\bp) &= h_1(\bp)  \nonumber \\
 + \frac{1}{2} \int \!\dbar^3 q
 & V_{\textsc{C},\textsc{F}}(\bk) \Bigg\{ \frac{B(\bp) B(\bq)
P_J + A(\bp)A(\bq)
\left( \frac{J}{2J+1} P_{J+1} + \frac{J+1}{2J+1} P_{J-1} \right)}{\omega(\bp) \omega(\bq)} g_1(\bq)  
 \nonumber \\ & \quad \quad 
+  \frac{A(\bp)}{\omega(\bp)} \frac{\sqrt{J
(J+1)}}{2J+1} \left( P_{J+1} -
P_{J-1} \right) g_2(\bq) \Bigg\}  \nonumber \\
+ \frac{1}{2} \int \!\dbar^3 q & 2 D(\bk) \Bigg\{ \frac{B(\bp)
B(\bq) \frac{H}{J} + A(\bp)A(\bq)
\left( \frac{J}{2J+1} F_2 + \frac{J+1}{2J+1}G_2 \right)}{\omega(\bp) \omega(\bq)} g_1(\bq)  \nonumber \\
& \quad \quad - \frac{A(\bq)}{\omega(\bq)} \frac{\sqrt{J
(J+1)}}{2J+1} \left( F_1 - G_1 \right)
g_2(\bq)
\Bigg\} , \;  \label{Mes2-inf2} \\ 
\omega_{\textsc{f}}(\bp) h_2(\bp) &=  \frac{\mu^2}{4} g_2(\bp) 
 \nonumber \\
+ \frac{1}{2} \int \!\dbar^3 q & V_{\textsc{C},\textsc{F}}(\bk)
\Bigg\{ \frac{B(\bp) B(\bq) P_J +
A(\bp)A(\bq) \left( \frac{J+1}{2J+1} P_{J+1} + \frac{J}{2J+1} P_{J-1}
\right)}{\omega(\bp) \omega(\bq)}
h_2(\bq)    \nonumber \\
& \quad \quad + \frac{A(\bq)}{\omega(\bq)} \frac{\sqrt{J
(J+1)}}{2J+1} \left( P_{J+1} -
P_{J-1}\right)
h_1(\bq)  \Bigg\}  \nonumber \\
+ \frac{1}{2} \int \!\dbar^3 q & 2 D(\bk) \Bigg\{ \frac{B(\bp)
B(\bq) P_J+ A(\bp)A(\bq)
\left( \frac{J+1}{2J+1} F_1 + \frac{J}{2J+1} G_1 \right)}{\omega(\bp) \omega(\bq)} h_2(\bq)  \nonumber \\
& \quad \quad - \frac{A(\bq)}{\omega(\bq)} \frac{\sqrt{J
(J+1)}}{2J+1} \left( F_2 - G_2 \right)
h_1(\bq)  \Bigg\} \; , \label{last-1} \\
\omega_{\textsc{f}}(\bp) g_2(\bq) &= h_2(\bp)  \nonumber
\\ + \frac{1}{2} \int \!\dbar^3 q & V_{\textsc{C},\textsc{f}}(\bk)
\Bigg\{ \left( \frac{J+1}{2J+1} P_{J+1} + \frac{J}{2J+1} P_{J-1}\right) g_2(\bq) +
 \frac{A(\bq)}{\omega(\bq)} \frac{\sqrt{J (J+1)}}{2J+1} \left(
P_{J+1} - P_{J-1}  \right)
g_1(\bq) \Bigg\}  \nonumber \\
-  \frac{1}{2} \int \!\dbar^3 q & 2 D(\bk) \Bigg\{ \left( \frac{J+1}{2J+1} F_1 + \frac{J}{2J+1} G_1 \right) g_2(\bq) 
+ \frac{A(\bq)}{\omega(\bq)} \frac{\sqrt{J (J+1)}}{2J+1}
\left(F_2 - G_2  \right) g_1(\bq)
\Bigg\} \; , \label{last-2}
\end{align}
\end{subequations}
with the quantities $F_1, F_2$,
Eqs.~(\ref{Def-F1}), (\ref{Def-G1}),
\begin{align}
\label{Def-F2} F_2 &= \frac{|\bp|}{|\bk|^2} \Big[ |\bp|
\left( P_{J-1}(\hat{\bp} \cdot \hat{\bq}) -  (\hat{\bp} \cdot
\hat{\bq}) P_{J}(\hat{\bp} \cdot \hat{\bq}) \right)  + |\bq|
\left( (\hat{\bp} \cdot  \hat{\bq}) P_{J+1}(\hat{\bp} \cdot
\hat{\bq})
- P_{J}(\hat{\bp} \cdot  \hat{\bq})  \right) \Big] \; , \\
\label{Def-G2} G_2 &= \frac{|\bp|}{|\bk|^2} \Big[
|\bp| \left( P_{J+1}(\hat{\bp} \cdot \hat{\bq}) -  (\hat{\bp}
\cdot \hat{\bq}) P_{J}(\hat{\bp} \cdot \hat{\bq}) \right)  +
 |\bq| \left(  (\hat{\bp} \cdot
\hat{\bq}) P_{J-1}(\hat{\bp} \cdot \hat{\bq}) -  P_{J}(\hat{\bp}
\cdot \hat{\bq}) \right) \Big] \; ,
\end{align}
and
\begin{align}
\label{H}
 H = \frac{|\bp| |\bq|}{|\bk|^2}  \left( (\hat{\bp} \cdot \hat{\bq}) P_{J}(\hat{\bp} \cdot \hat{\bq}) - P_{J+1}(\hat{\bp} \cdot \hat{\bq}) \right) \; ,
\end{align}
entering the transverse gluon part. We note that for the $0^{++}$ state several vertex function components in Eq.~(\ref{Cat2}) are absent and therefore
only two independent functions $h_2(\bp)$ and $g_2(\bp)$ remain. 

Finally, let us turn to the Bethe--Salpeter equations for mesons of
category three (\ref{Mes3-1}), (\ref{Mes3-2}), which for $\mu_{\textsc{IR}}\rightarrow 0$ have the form 
\begin{subequations}
\begin{align}
\label{Mes3-inf1}
\omega_{\textsc{F}}(\bp) g(\bp) &= h(\bp) +
 \frac{1}{2} \int \dbar^3 q \, \left( V_{\textsc{C}}(\bk) P_{J}(\hat{\bp} \cdot \hat{\bq}) +
 2  D(\bk) \frac{H}{J} \right)  g(\bq) \, , \\
 \label{Mes3-inf2}
 \omega_{\textsc{F}}(\bp) h(\bp) &= \frac{\mu^2}{4} g(\bp)  \nonumber \\
 + \frac{1}{2} \int \dbar^3 q & V_{\textsc{C}}(\bk) \Bigg\{
\frac{A(\bp) A(\bq) P_{J}(\hat{\bp} \cdot \hat{\bq}) + B(\bp)
B(\bq) \left(\frac{J}{2J+1} P_{J+1}(\hat{\bp} \cdot \hat{\bq})  +
\frac{J+1}{2J+1} P_{J-1}(\hat{\bp} \cdot \hat{\bq})
\right)}{\omega(\bp) \omega(\bq)} \Bigg\}
 h(\bq)   \nonumber \\
- \frac{1}{2} \int \dbar^3 q & \, 2 \, D(\bk)\Bigg\{ \frac{A(\bp) A(\bq)
\frac{H}{J} + B(\bp) B(\bq) \left( \frac{J}{2J+1} F_2
+ \frac{J+1}{2J+1} G_2 \right)}{\omega(\bp)
\omega(\bq)} h(\bq) \Bigg\} \; ,
\end{align}
\end{subequations}
where we have used the definitions 
(\ref{Def-F2}), (\ref{Def-G2}), (\ref{H}). For the state $0^{--}$ there would only be one independent vertex tensor component, see Eq.~(\ref{Cat3}). 
However, inserting it into the BSE gives zero. Hence, such a state is absent in our instantaneous model, Ref.~\cite{Wagenbrunn:2007ie}. 

All equations listed in this section now serve as
a laboratory to study the influence of transverse gluons on mesons for large orbital excitations. 
By analyzing the occurring angular integrals we are going to observe that the transverse gluon interaction is suppressed for highly excited
meson states. 

\section{Chiral Symmetry Structure of the Bethe--Salpeter Equations\\ for the Non-Interacting Case}
\label{Mass-zero-chap}
In this section we discuss the special case of non-interacting quarks and show that the BSEs fulfill
the expected meson degeneracies, Eqs.~(\ref{mes-deg})-(\ref{mes-deg3}).
For vanishing dynamical quark mass $M(\bp)=0$ the integral equations (\ref{Mes1-inf1}),
(\ref{Mes1-inf2}) for mesons of category one simplify as
(using Eq.~(\ref{tree-level}) and $\omega(\bp) = |\bp|$)
\begin{subequations}
\begin{align}
\label{Mes1-tree1}
 \omega_{\textsc{F}}(\bp)  h(\bp) &=
\frac{\mu^2}{4} g(\bp) + \frac{1}{2} \int \dbar^3 q \,
\left(V_{\textsc{C}, \textsc{F}}(\bk) + 2 D(\bk) \right)
P_{J}(\hat{\bp} \cdot \hat{\bq}) h(\bq) \; , \\
\label{Mes1-tree2}
 \omega_{\textsc{F}}(\bp) g(\bp) &= h(\bp)
+ \frac{1}{2} \int \dbar^3 q V_{\textsc{C},\textsc{F}}(\bk)
\left\{
\frac{J+1}{2J+1} P_{J+1}(\hat{\bp} \cdot \hat{\bq})  + \frac{J}{2J+1} P_{J-1}(\hat{\bp} \cdot \hat{\bq}) \right\} g(\bq)  
\nonumber \\ & \quad \quad \quad - 
 \frac{1}{2} \int \dbar^3 q \, 2 \, D(\bk)
\left\{ \frac{J+1}{2J+1} F_1  + \frac{J}{2J+1}
G_1 \right\} g(\bq) \; ,
\end{align}
\end{subequations}
with $F_1$ and $G_1$ given in Eqs.~(\ref{Def-F1}), (\ref{Def-G1}).
Inserting the tree-level dressing
functions into the last two equations for mesons of category two, (\ref{last-1}), (\ref{last-2}) yields
\begin{subequations}
\begin{align}
\label{Mes2-tree31}
\omega_{\textsc{F}}(\bp) h_2(\bp) &=  \frac{\mu^2}{4} g_2(\bp) +
\frac{1}{2} \int \!\dbar^3 q \left( V_{\textsc{C},\textsc{F}}(\bk) + 2 D(\bk) \right)
  P_J(\hat{\bp} \cdot \hat{\bq})
h_2(\bq)    \; , \\
\label{Mes2-tree32}
\omega_{\textsc{f}}(\bp) g_2(\bq) & = h_2(\bp) + \frac{1}{2} \int \!\dbar^3 q  V_{\textsc{C},\textsc{f}}(\bk)
 \left( \frac{J+1}{2J+1} P_{J+1}(\hat{\bp} \cdot \hat{\bq}) + \frac{J}{2J+1} P_{J-1}(\hat{\bp} \cdot \hat{\bq}) \right) g_2(\bq)  \nonumber \\
& \quad \quad \quad -  \frac{1}{2} \int \!\dbar^3 q \, 2 D(\bk)
\left( \frac{J+1}{2J+1} F_1 + \frac{J}{2J+1}
G_1 \right) g_2(\bq) \; ,
\end{align}
\end{subequations}
and one ends up with identical equations, which yield the expected degeneracies $J^{-+} \longleftrightarrow J^{++} (J=2n)$ and $J^{+-} \longleftrightarrow J^{--} (J=2n+1)$.
The remaining two equations (\ref{Mes2-inf1}),
(\ref{Mes2-inf2}) for mesons of category two 
\begin{subequations}
\begin{align}
\label{Mes2-vanM1} \omega_{\textsc{f}}(\bp) h_1(\bp) &=
\frac{\mu^2}{4} g_1(\bp) + \frac{1}{2} \int \dbar^3 q \,
V_{\textsc{C},\textsc{f}}(\bk) \Bigg\{ \left( \frac{J}{2J+1}
P_{J+1}(\hat{\bp} \cdot \hat{\bq})
+ \frac{J+1}{2J+1} P_{J-1}(\hat{\bp} \cdot \hat{\bq}) \right) h_1(\bq)  \nonumber \\
& \quad \quad \quad -  \frac{1}{2}  \int \dbar^3 q  \, 2 D(\bk)
\Bigg\{ \left( \frac{J}{2J+1} F_2
+ \frac{J+1}{2J+1} G_2 \right)  h_1(\bq) \; , \\
\label{Mes2-vanM2}
 \omega_{\textsc{f}}(\bp)
 g_1(\bp) &= h_1(\bp) + \! \frac{1}{2} \int \!\dbar^3 q
\, \left( V_{\textsc{C},\textsc{F}}(\bk) P_J(\hat{\bp} \cdot
\hat{\bq}) +  2 D(\bk) \frac{H}{J} \right) g_1(\bq)  \; ,
\end{align}
\end{subequations}
then coincide with the equations for mesons of category three:
\begin{subequations}
\begin{align}
\omega_{\textsc{F}}(\bp) h(\bp) &= \frac{\mu^2}{4} g(\bp) 
 + \frac{1}{2} \int \dbar^3 q  V_{\textsc{C},\textsc{f}}(\bk)
 \left(\frac{J}{2J+1} P_{J+1}(\hat{\bp} \cdot \hat{\bq})  +
\frac{J+1}{2J+1} P_{J-1}(\hat{\bp} \cdot \hat{\bq}) \right)
 h(\bq)   \nonumber \\
& \quad \quad \quad - \frac{1}{2} \int \dbar^3 q  \, 2 \, D(\bk)  \left(
\frac{J}{2J+1} F_2 + \frac{J+1}{2J+1}
G_2 \right) h(\bq)  \, , \\ 
\omega_{\textsc{F}}(\bp) g(\bp) &= h(\bp) +
 \frac{1}{2} \int \dbar^3 q \, \left( V_{\textsc{C},\textsc{f}}(\bk) P_{J}(\hat{\bp} \cdot \hat{\bq}) +
 2  D(\bk) \frac{H}{J} \right)  g(\bq) \; .
\end{align}
\end{subequations}
The meson degeneracies $J^{--} \longleftrightarrow J^{++} (J=2n+1)$ and $J^{++} \longleftrightarrow J^{--} (J=2(n+1))$
follow from these equations. The state $0^{++}$ only applies for the last two equations of category two, Eqs.~(\ref{Mes2-tree31}), (\ref{Mes2-tree32}), 
as shown in the last section.  
We have thus shown, that for vanishing dynamical mass the BSEs fall into the
expected chiral multiplets and $ SU_{\textsc{L}}(2) \times SU_{\textsc{R}}(2)$ and $U_{\textsc{A}}(1)$ are recovered. 
Next we analyze this question for the interacting case
and discuss the role played by transverse gluons. 

\section{Effective Chiral Symmetry Restoration for Large Orbital Angular Momentum}
\label{Restor}
The transverse gluon interaction influences the meson states in two ways, namely via the BSEs and via the quark dressing functions.
In this section we analyze the BSEs in order to answer if transverse gluons change the asymptotic degeneracy for high-spin bound states. 

By comparing mesons of category one and three, Eqs.~(\ref{Mes1-inf1}), (\ref{Mes1-inf2}) and  (\ref{Mes3-inf1}), (\ref{Mes3-inf2}), 
in the limit of large spins $J$ for a purely linear interquark potential $V_{\textsc{C}}$, 
the degeneracy of all chiral multiplets, Eq.~(\ref{chir-mul}), 
has been demonstrated in Refs.~\cite{Wagenbrunn:2006cs, Wagenbrunn:2007ie}. 
Here we discuss the effect of the additional transverse gluon interaction, Eq.~(\ref{Ham-trans}),
for the degeneracy within these two categories in the limit $J\rightarrow \infty$.
For mesons in category three, Eqs.~(\ref{Mes3-inf1}), (\ref{Mes3-inf2}), we make the
replacement $h(\bp) \rightarrow \frac{\mu}{2} g(\bp)$. For a clear reading we once again list the equations for category one
\begin{subequations}
\begin{align}
\label{Mes1-inf1-B}
 \omega_{\textsc{F}}(\bp)  h(\bp) &=
\frac{\mu^2}{4} g(\bp) +  \frac{1}{2} \int \dbar^3 q \,
\left(V_{\textsc{C}, \textsc{F}}(\bk) + 2 D(\bk) \right)
P_{J}(\hat{\bp} \cdot \hat{\bq}) h(\bq) \; , \\
\label{Mes1-inf2-B}
\omega_{\textsc{F}}(\bp) g(\bp) &= h(\bp)  \nonumber \\
+ \frac{1}{2} \int \dbar^3 q V_{\textsc{C},\textsc{F}}(\bk)
&\left\{ \frac{A(\bp) A(\bq) P_{J}(\hat{\bp} \cdot \hat{\bq}) +
B(\bp) B(\bq)
\left(\frac{J+1}{2J+1} P_{J+1}(\hat{\bp} \cdot \hat{\bq})  + \frac{J}{2J+1} P_{J-1}(\hat{\bp} \cdot \hat{\bq}) \right)}{\omega(\bp) \omega(\bq)} \right\} g(\bq) \nonumber \\
- \frac{1}{2} \int \dbar^3 q \, 2 \, D(\bk) & \left\{
\frac{A(\bp)A(\bq) P_{J}(\hat{\bp} \cdot \hat{\bq}) + B(\bp)B(\bq)
\left(\frac{J+1}{2J+1} F_1  + \frac{J}{2J+1}
G_1 \right)}{\omega(\bp) \omega(\bq)} \right\} g(\bq) \; ,
\end{align}
\end{subequations}
and for category three
\begin{subequations}
\begin{align}
\label{Mes3-inf1-B}
\omega_{\textsc{F}}(\bp) h(\bp) &= \frac{\mu^2}{4} g(\bp) +
 \frac{1}{2} \int \dbar^3 q \, \left( V_{\textsc{C},\textsc{F}}(\bk) P_{J}(\hat{\bp} \cdot \hat{\bq}) +
 2  D(\bk) \frac{H}{J} \right)  h(\bq) \, , \\
\label{Mes3-inf2-B}
\omega_{\textsc{F}}(\bp) g(\bp) &= h(\bp)  \nonumber \\
 + \frac{1}{2} \int \dbar^3 q V_{\textsc{C}}(\bk) & \left\{
\frac{A(\bp) A(\bq) P_{J}(\hat{\bp} \cdot \hat{\bq}) + B(\bp)
B(\bq) \left(\frac{J}{2J+1} P_{J+1}(\hat{\bp} \cdot \hat{\bq})  +
\frac{J+1}{2J+1} P_{J-1}(\hat{\bp} \cdot \hat{\bq})
\right)}{\omega(\bp) \omega(\bq)}
\right\}  g(\bq)  \nonumber \\
- \frac{1}{2} \int \dbar^3 q  \, 2 \, D(\bk) & \left\{\frac{A(\bp) A(\bq)
\frac{H}{J} + B(\bp) B(\bq) \left( \frac{J}{2J+1} F_2
+ \frac{J+1}{2J+1} G_2 \right)}{\omega(\bp)
\omega(\bq)} \right\}  g(\bq)  \; ,
\end{align}
\end{subequations}
using the definitions (\ref{Def-F1}), (\ref{Def-G1}), (\ref{Def-F2}), (\ref{Def-G2}), (\ref{H}).
Let us start with exploring Eqs.~(\ref{Mes1-inf1-B}) and (\ref{Mes3-inf1-B}).
The color-Coulomb interaction $V_{\textsc{C}}(\bk)$ enters in both equations 
with the Legendre Polynomial $P_{J}(\hat{\bp} \cdot \hat{\bq})$, hence for any spin quantum number $J$
the equations agree with each other.
However, for the transverse gluon interaction the situation is different: two different
contributions $P_{J}(\hat{\bp} \cdot \hat{\bq})$ and
$\frac{H}{J}$ occur, which do not coincide for arbitrary $J$. However,
we now demonstrate that for large spin quantum number $J$ both these
contributions vanish.

\begin{figure}[t]
 \centering
 \includegraphics[angle=0,width=.45\linewidth]{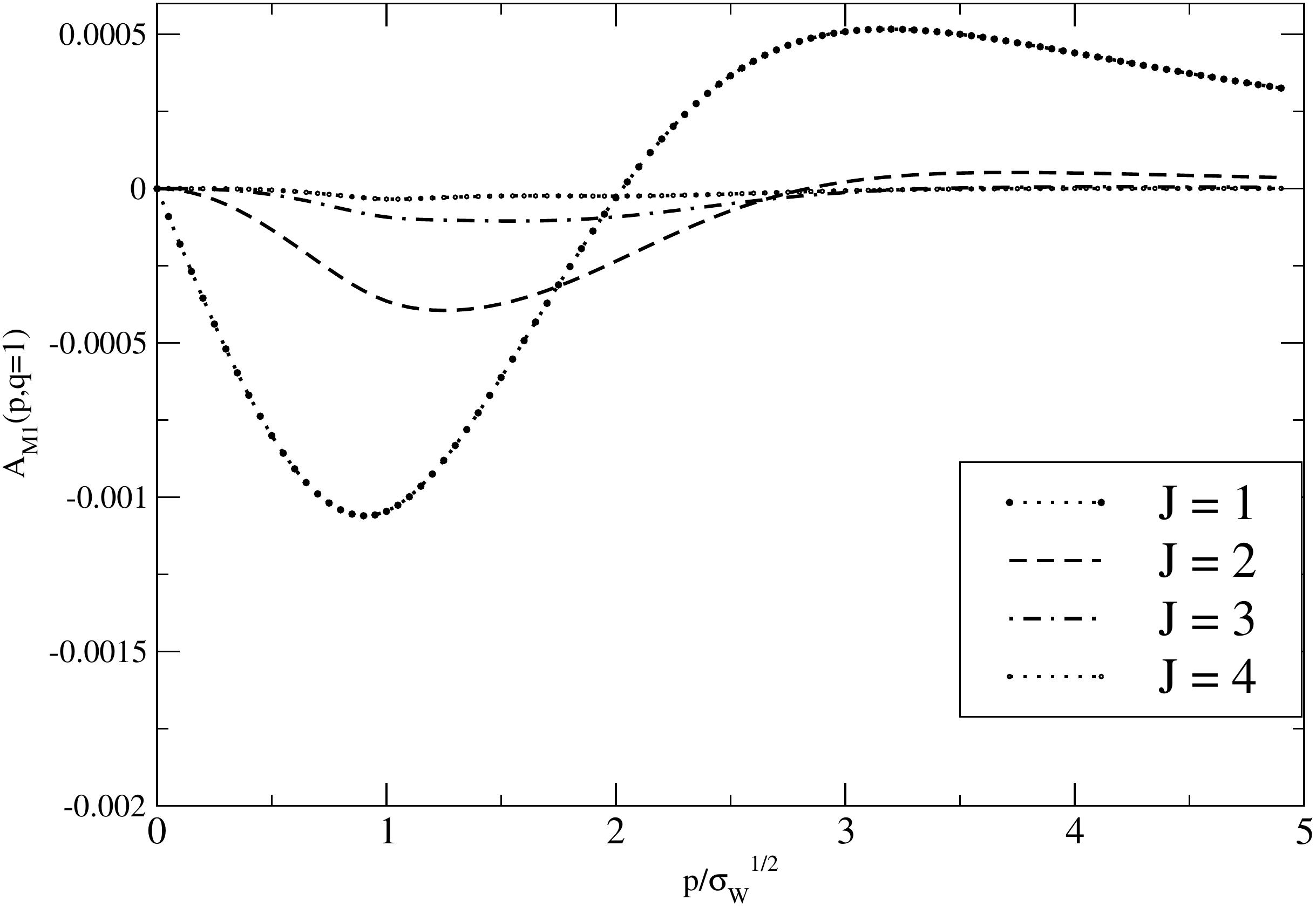}
 \includegraphics[angle=0,width=.45\linewidth]{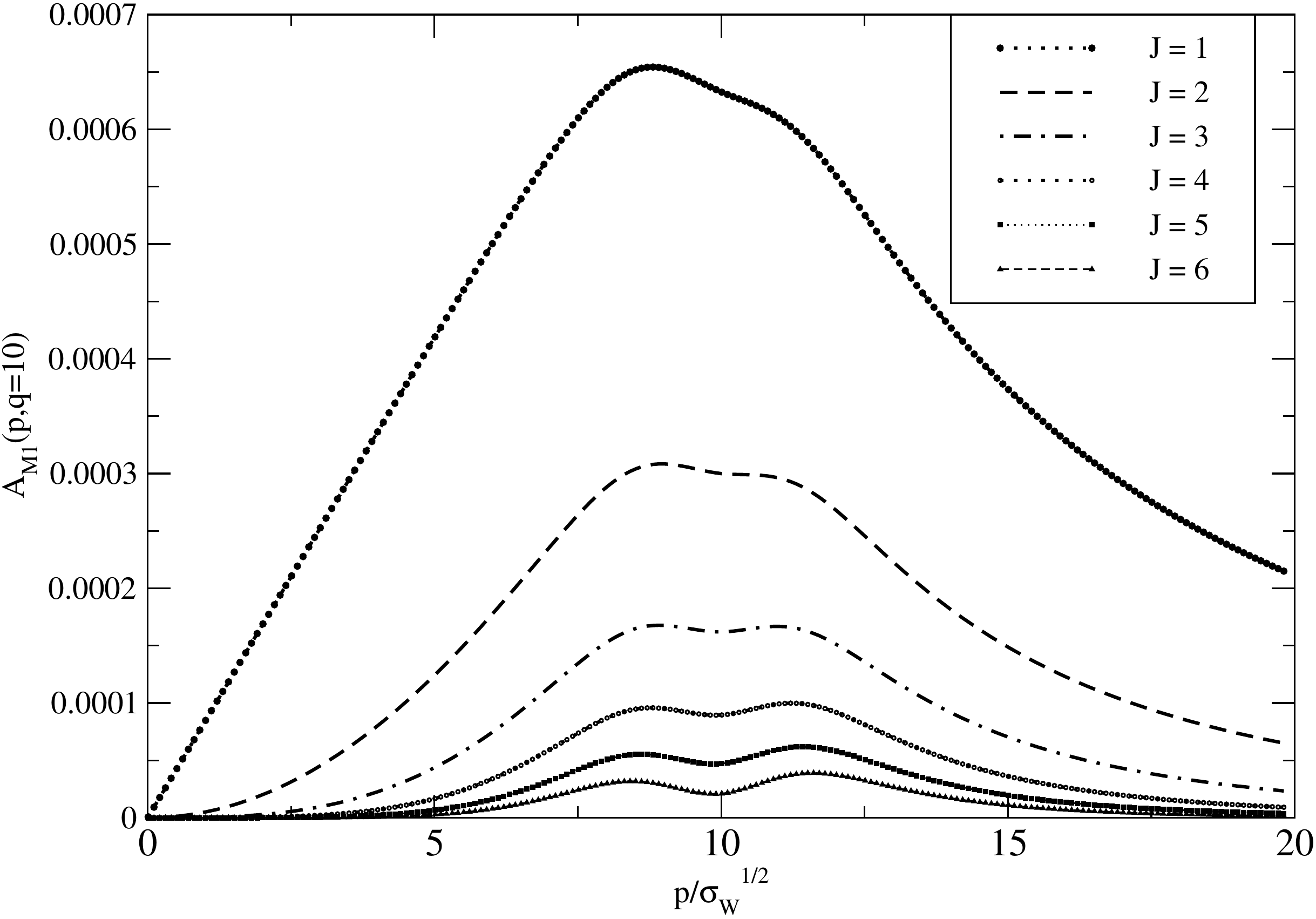}
 \caption[Angular integral I.]{\sl Integral (\ref{I-Integral}) for different values of spin $J$ and $q/\sqrt{\sigma_{\textsc{W}}}$ fixed at $1,10$.}
 \label{fig-IJ-2}
 \end{figure}

\begin{figure}[t]
 \centering
 \includegraphics[angle=0,width=.45\linewidth]{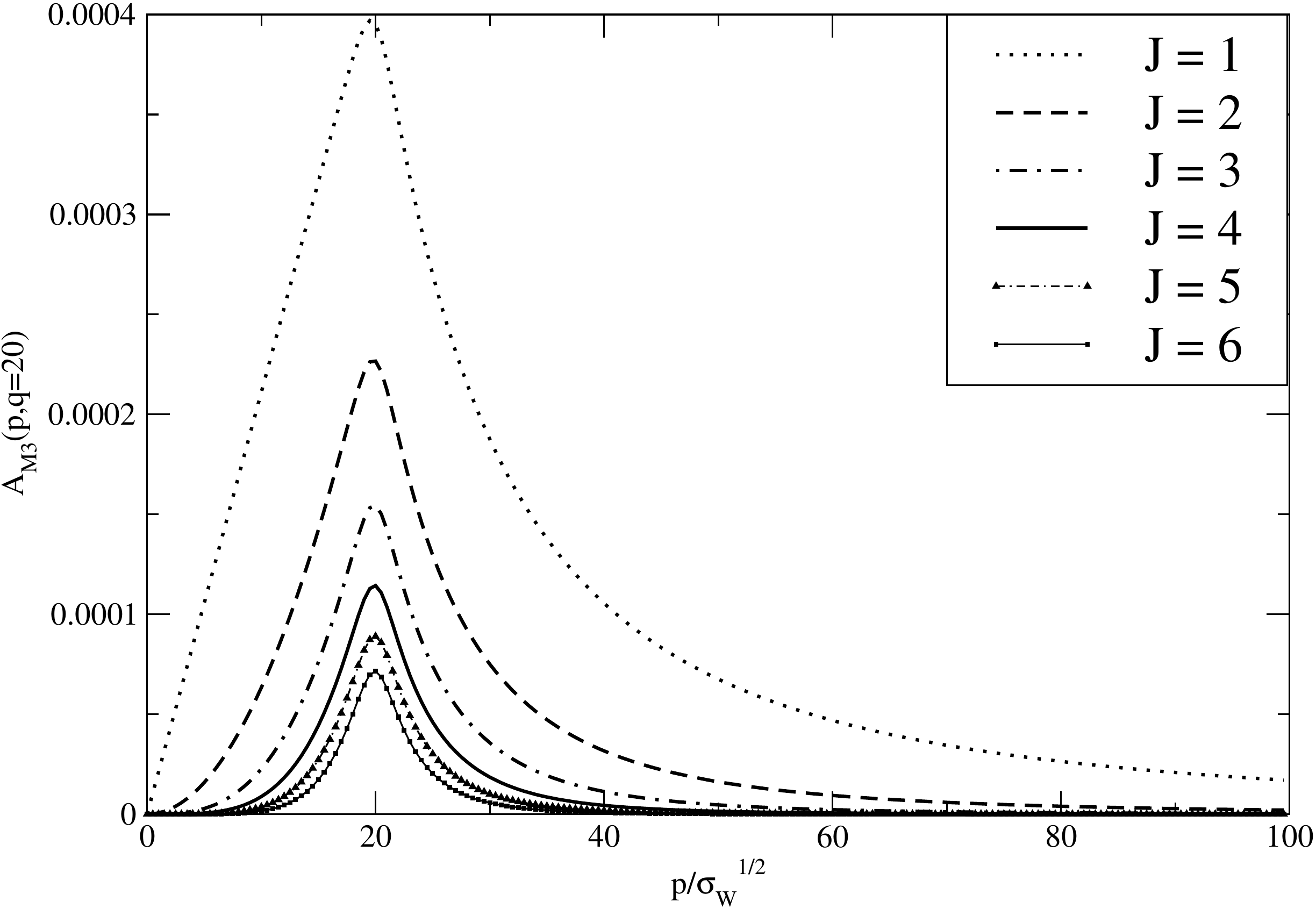}
 \includegraphics[angle=0,width=.45\linewidth]{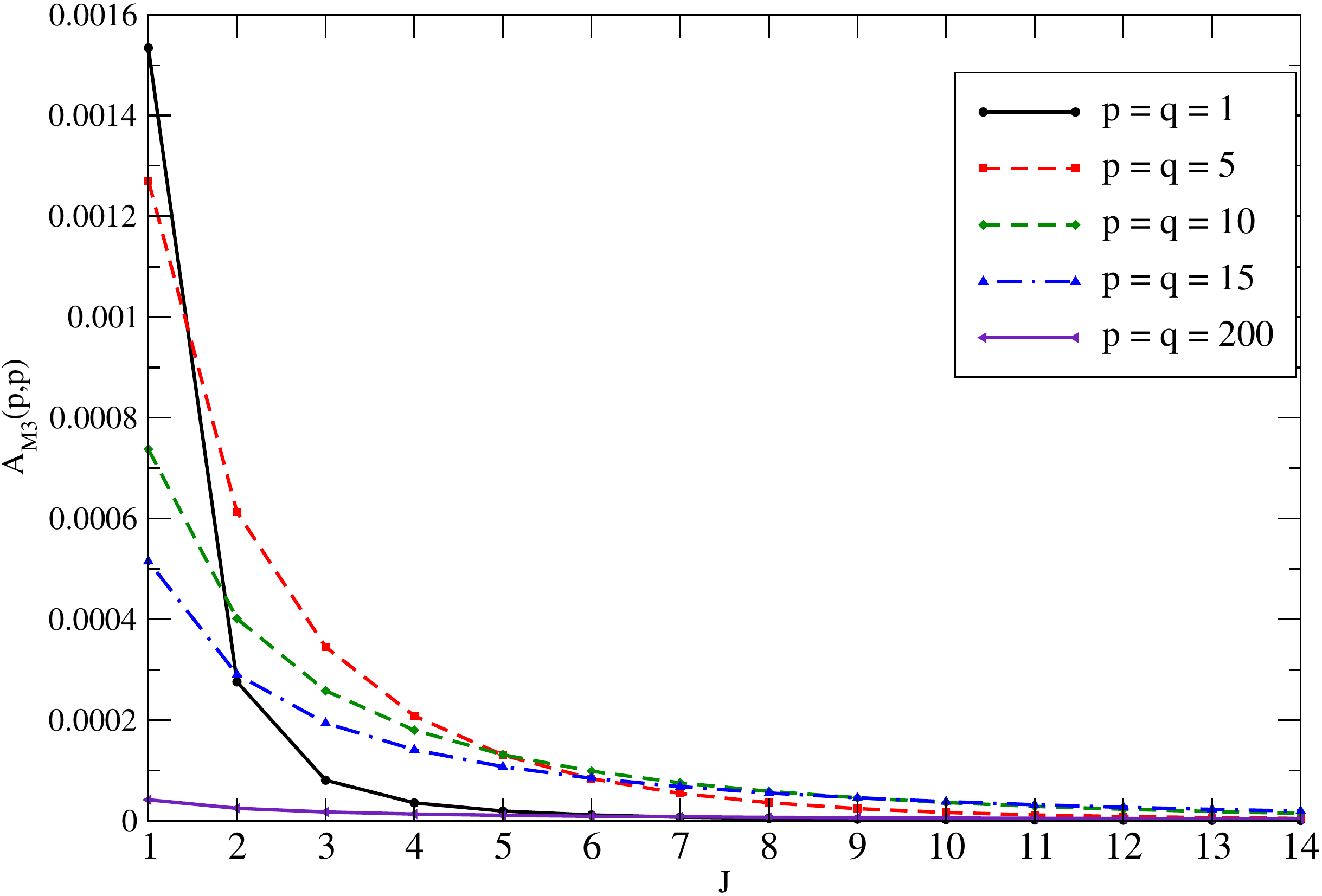}
 \caption[Angular integral I.]{\sl \textit{Left Panel:} Integral (\ref{J-Integral}) for different values of spin $J$ and fixed $q/\sqrt{\sigma_{\textsc{W}}}=20$.
\textit{Right Panel:} The same integral at the maximum ($p=q$) for different values of spin $J$. }
 \label{fig-J}
 \end{figure}

We use spherical coordinates ($\int d^3 q = 
2\pi \int_{0}^{\infty} q^2 dq \int_{-1}^{1} z $, 
with $z=\cos\theta = \hat{\bp} \cdot \hat{\bq}$) and write down the one-dimensional integral equations
\begin{align}
\label{one-dim1}
\omega_{\textsc{F}}(p)  h(p) &=
\frac{\mu^2}{4} g(p) +  \frac{1}{2} \frac{1}{(2\pi)^2} \int q^2 dq \, h(q) \int_{-1}^{1} dz \,
\left(V_{\textsc{C}, \textsc{F}}(p^2+q^2-2pqz) + 2 D(p^2+q^2-2pqz) \right)
P_{J}(z)  \; , \\ 
\label{one-dim2}
\omega_{\textsc{F}}(p) h(p) &= \frac{\mu^2}{4} g(p) +
 \frac{1}{2} \frac{1}{(2\pi)^2} \int q^2 dq \, h(q) \, \int_{-1}^{1} dz \left( V_{\textsc{C},\textsc{F}}(p^2+q^2-2pqz) P_{J}(z) +
 2  D(p^2+q^2-2pqz) \frac{H(p,q,z)}{J} \right) \, .
\end{align}
We calculate the angular integrals for the transverse gluon interaction
\begin{align}
\label{I-Integral}
A_{\textsc{M1}}(p,q) &= \frac{1}{(2\pi)^2} \int_{-1}^{1} dz \, D(p^2+q^2-2pqz)
P_{J}(z) \; , \\
\label{J-Integral}
A_{\textsc{M3}}(p,q) &= \frac{1}{(2\pi)^2} \frac{1}{J} \int_{-1}^{1} dz
D(p^2+q^2-2pqz) \, H(p,q,z)  
 = \frac{1}{(2\pi)^2} \frac{p q}{J} \left\{ \int_{-1}^{1} dz
\frac{D(p^2+q^2-2pqz)}{p^2+q^2-2pqz} \left( z P_{J}(z) -  P_{J+1}(z) \right) \right\} ,
\end{align}
where in the second line we have explicitly used the definition for $H$, Eq.~(\ref{H}).
For a tree-level gluon propagator the integrals
can be calculated analytically \footnote{Using a tree-level gluon propagator 
$$
 D(\bk) \rightarrow D_{\textsc{UV}}(\bk)= \frac{1}{2 \, |\bk|} \; ,
$$
 the angular integrals (\ref{I-Integral}), (\ref{J-Integral}) can be analytically performed.
The integral (\ref{I-Integral}) then simplifies as
$$
 A_{\textsc{UV}, \textsc{M1}}(p,q) = \frac{1}{(2\pi)^2} \frac{1}{2J+1} \left(\left(\frac{q}{p} \right)^{J} \frac{1}{p} \theta(p-q) + \left(\frac{p}{q} \right)^{J} \frac{1}{q} \theta(q-p) \right) \; .
$$
The maximum of the function $I_{\textsc{UV}}(p,p=q,J)$ vanishes for increasing $J$ as $\frac{1}{2J+1} \frac{1}{p}$. For the angular integral (\ref{J-Integral}) the extremum at $p=q$ 
can also be shown to vanish as 
$$
 A_{\textsc{UV}, \textsc{M3}}(p) \sim \frac{1}{2J+1} \frac{1}{p} \; . 
$$
It is interesting to note,
that the angular integral (\ref{J-Integral}) for the individual contributions $z P_J$ and $P_{J+1}$ does not converge, only the difference gives a well-defined integral for the tree-level 
gluon propagator.}.
Using as transverse gluon propagator $D$ the lattice result, 
Eq.~(\ref{transv-gluon}) the angular integrals (\ref{I-Integral}), (\ref{J-Integral}) are performed numerically. 
The scale is fixed by the Wilsonian string tension $\sqrt{\sigma_{\textsc{W}}} =440$ MeV. The fact, that the first integral, Eq.~(\ref{I-Integral}), vanishes for $J\rightarrow \infty$ 
is not surprising and comes from the properties of the Legendre Polynomials. 
In Fig.~\ref{fig-IJ-2} we plot this integral for fixed values of $q$. It can be seen, that the function values fast become smaller for larger values of $J$.
The same calculation is performed for Eq.~(\ref{J-Integral}), see Fig.~\ref{fig-J}. 
Again, the contributions decrease as $J$ is increased. On the right-hand side of Fig.~\ref{fig-J} the values of the maximum ($p=q$) are 
evaluated for different orbital quantum numbers $J$. Moreover, going to larger values for the outer momentum $p$ gives smaller values for given $J$. 
We can conclude, that for large $J$
both these integrals (\ref{I-Integral}), (\ref{J-Integral}) vanish and the integral equations (\ref{one-dim1}), (\ref{one-dim2}) 
are in the limit $J \rightarrow \infty$ completely determined by the color-Coulomb interaction $V_{\textsc{C}}$, 
Eq.~(\ref{Ham-Coul}).  

Let us turn to the second system of equations, (\ref{Mes1-inf2-B}) and (\ref{Mes3-inf2-B}).
For the color-Coulomb potential part, the term on the right-hand side with the scalar quark dressing function $A(\bp)$ coincides for both equations. 
In the terms which enter with the vector dressing function $B(\bp)$ we get the contributions
\begin{align}
 \frac{J+1}{2J+1} P_{J+1}(\hat{\bp} \cdot \hat{\bq})  + \frac{J}{2J+1} P_{J-1}(\hat{\bp} \cdot \hat{\bq}) \; , 
\end{align}
and
\begin{align}
 \frac{J}{2J+1} P_{J+1}(\hat{\bp} \cdot \hat{\bq})  + \frac{J+1}{2J+1} P_{J-1}(\hat{\bp} \cdot \hat{\bq}) \; ,
\end{align}
which coincide for $J \rightarrow \infty$, as already outlined in Ref.~\cite{Wagenbrunn:2007ie}. For the transverse gluon interaction part we get two contributions.
 For the
terms with the scalar dressing function $A(\bp)$  on the right-hand side of the equations (\ref{Mes1-inf2-B}) and (\ref{Mes3-inf2-B}) we have the contributions $P_J$ and
$\frac{1}{J} H$, which have already been shown to
vanish for large $J$. The remaining terms are 
\begin{align}
\frac{J+1}{2J+1} F_1  + \frac{J}{2J+1} 
G_1 \; , 
\end{align}
and
\begin{align}
\frac{J}{2J+1} F_2 + \frac{J+1}{2J+1}
G_2 \; . 
\end{align}
We write down the angular integrals of the two equations 
(\ref{Mes1-inf2-B}) and (\ref{Mes3-inf2-B}) for the transverse gluon part:
\begin{align}
B_{\textsc{M1}}(p,q) &= \frac{1}{(2\pi)^2} \frac{1}{2}\int_{-1}^{1} dz \, D(p^2+q^2-2pqz)
\frac{pz-q}{p^2+q^2-2pqz}
\Bigg\{ \frac{J+1}{2J+1} \Big[ p P_J(z) - q P_{J+1}(z)\Big]   \nonumber \\
\label{L-Integral2}
 & \quad \quad \quad \quad + \frac{J}{2J+1}
\Big[ p P_J(z) - q P_{J-1}(z) \Big] \Bigg\}
\; , \\
\label{M-Integral2}
B_{\textsc{M3}}(p,q) &= \frac{1}{(2\pi)^2} \frac{1}{2} \int_{-1}^{1} dz \, D(p^2+q^2-2pqz)
\frac{p}{p^2+q^2-2pqz} \Bigg\{
\frac{J}{2J+1} \Big[ p \left( P_{J-1}(z) - z P_{J}(z)\right) + q \left( z P_{J+1}(z) - P_J(z)\right) \Big]  
 \nonumber \\
& \quad \quad \quad \quad + \frac{J+1}{2J+1} \Big[p \left( P_{J+1}(z) - z P_{J}(z)\right) + q \left( z P_{J-1}(z) - P_J(z)\right) \Big]  \Bigg\} \; ,
\end{align}
using the definitions (\ref{Def-F1}), (\ref{Def-G1}), (\ref{Def-F2}), (\ref{Def-G2}).
 \begin{figure}[t]
 \centering
 \includegraphics[angle=0,width=.45\linewidth]{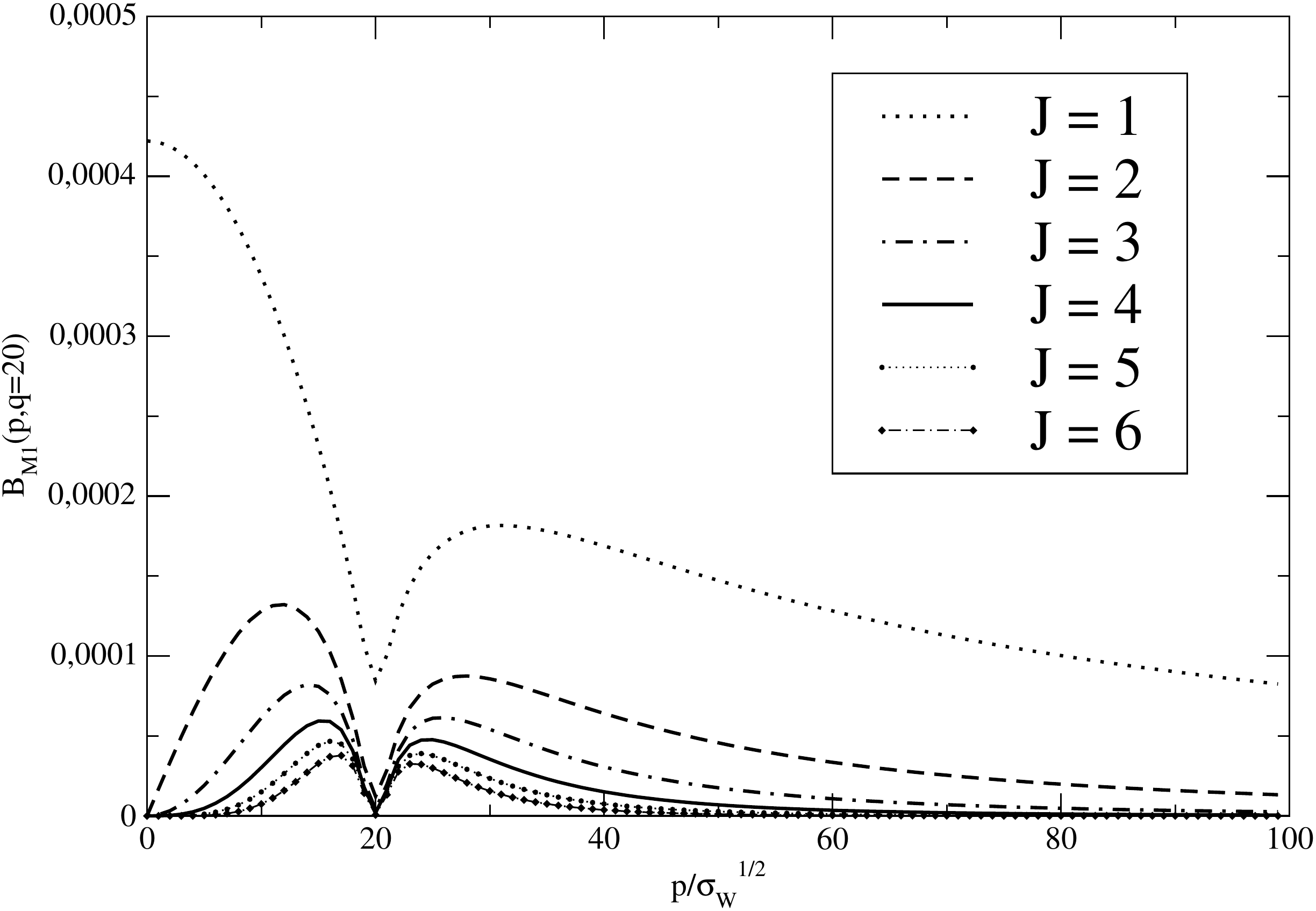}
 \includegraphics[angle=0,width=.45\linewidth]{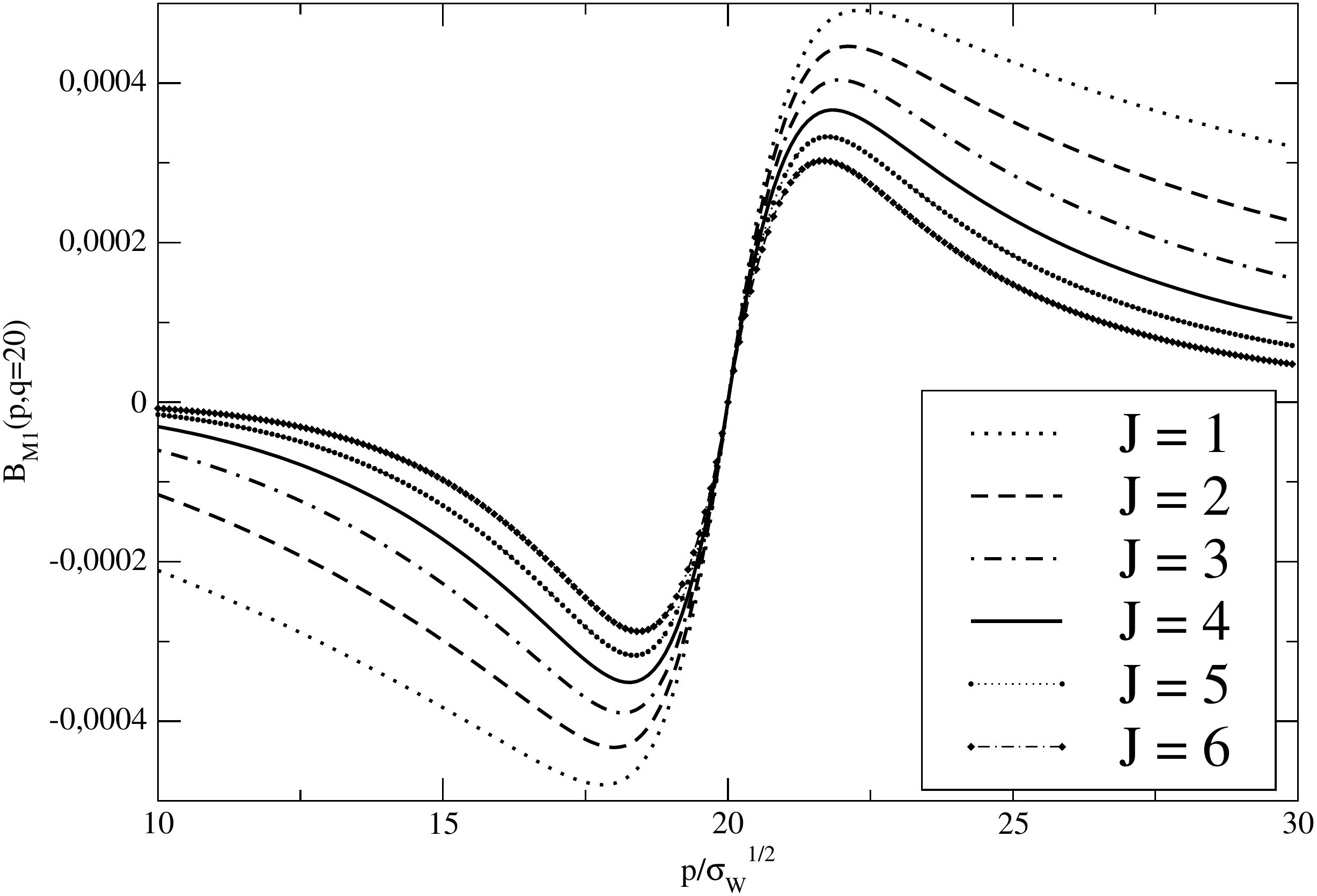}
 \caption[Angular integral I.]{\sl
 Integrals (\ref{L-Integral2}) (l.h.s) and (\ref{L-Int-part2}) (r.h.s) for different values of spin $J$ and fixed $q/\sqrt{\sigma_{\textsc{W}}}=20$.}
 \label{fig-KLM-2}
 \end{figure}
 \begin{figure}[t]
 \centering
 \includegraphics[angle=0,width=.45\linewidth]{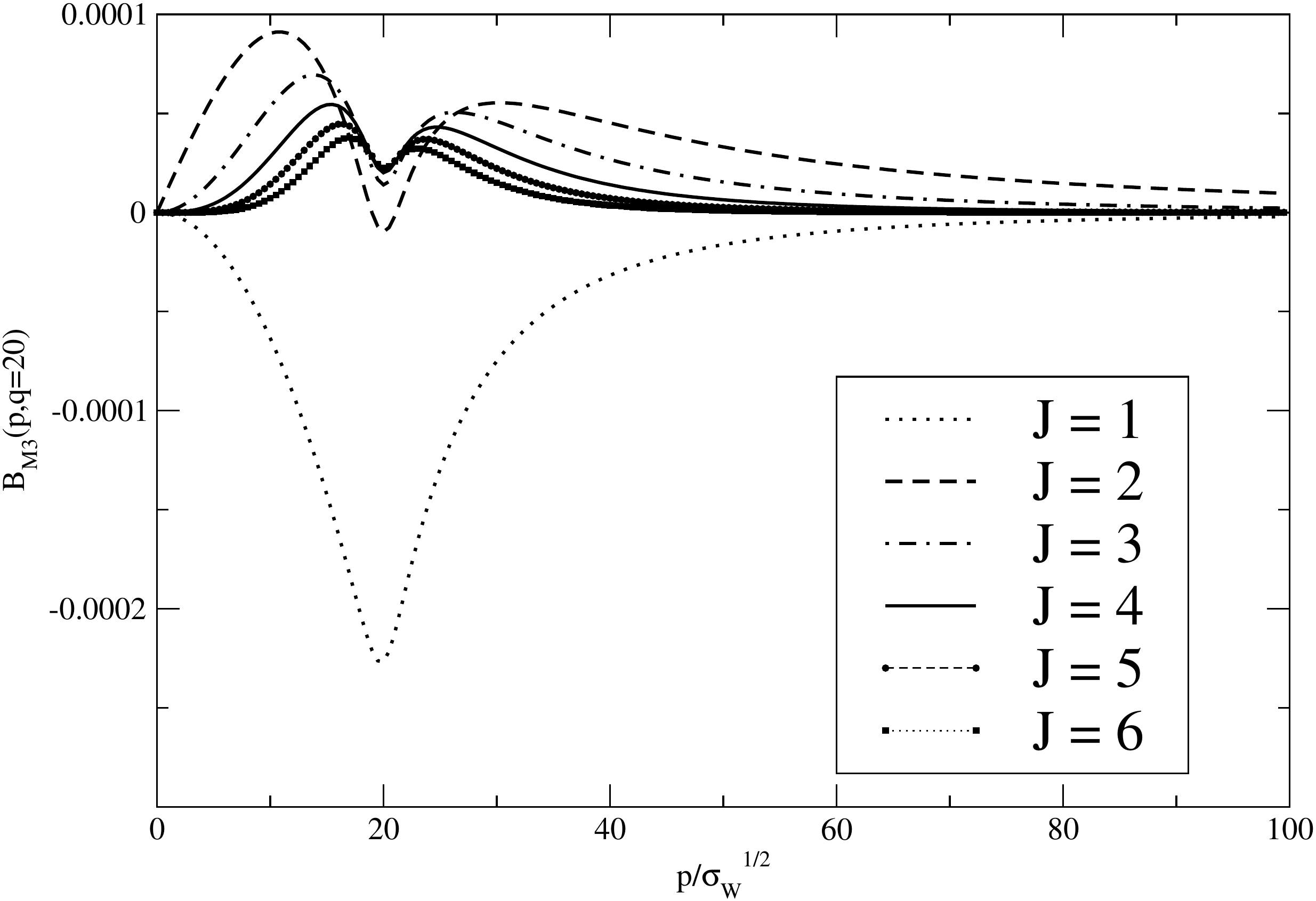}
 \includegraphics[angle=0,width=.45\linewidth]{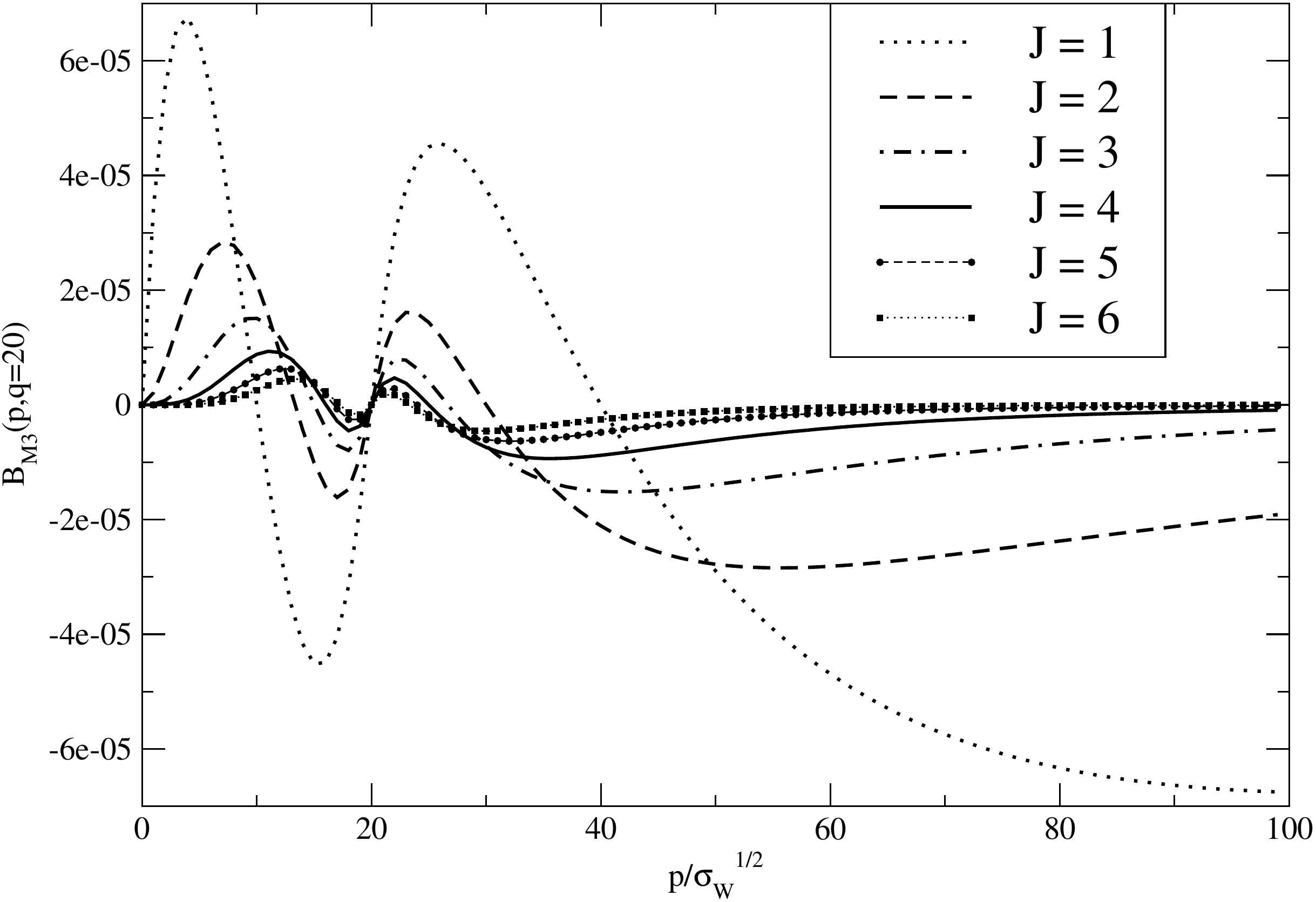}
 \caption[Angular integral I.]{\sl
 Integrals (\ref{M-Integral2}) (l.h.s) and (\ref{M-Int-part}) (r.h.s) for different values of spin $J$ and fixed $q/\sqrt{\sigma_{\textsc{W}}}=20$.}
 \label{fig-KLM-3}
 \end{figure}
The functions do not coincide for arbitrary spin quantum number $J$, however, they
vanish for large $J$.
In Fig.~(\ref{fig-KLM-2}) on the left-hand side $B_{\textsc{M1}}(p,q)$, Eq.~(\ref{L-Integral2}) is shown. It has a minimum at $p=q$ and the maxima
flatten for larger values of $J$. For $J\rightarrow \infty$ all Legendre Polynomials can be replaced by $P_{J+1}, P_{J-1} \rightarrow P_J$ and the integral (\ref{L-Integral2}) approaches 
\begin{align}
\label{L-Int-part}
 \frac{1}{(2\pi)^2} \frac{1}{2}\int_{-1}^{1} dz \, D(p^2+q^2-2pqz)
\frac{(pz-q) (p-q)}{p^2+q^2-2pqz} P_{J\rightarrow \infty}(z) \; . 
\end{align}
The decrease of $B_{\textsc{M1}}(p,q)$ for large $J$ can nicely be seen by plotting the individual integral contributions. The integral 
\begin{align}
\label{L-Int-part2}
\frac{1}{(2\pi)^2} \frac{1}{2}\int_{-1}^{1} dz \, D(p^2+q^2-2pqz)
\frac{pz (p-q)}{p^2+q^2-2pqz} P_{J\rightarrow \infty}(z) 
\end{align}
is shown on the right-hand side of Fig.~(\ref{fig-KLM-2}). For the function $B_{\textsc{M3}}(p,q)$, Eq.~ (\ref{M-Integral2}), the situation is analogous, 
see left-hand side of Fig.~\ref{fig-KLM-3}. 
It has a minimum at $p=q$, which goes to zero as $J$ is increased. In the case of $J \rightarrow \infty$ 
 the integral approaches
\begin{align}
\label{M-Int-part}
 \frac{1}{(2\pi)^2} \frac{1}{2} \int_{-1}^{1} dz \, D(p^2+q^2-2pqz)
\frac{p (p-q)}{p^2+q^2-2pqz} (1-z) \, P_{J\rightarrow \infty}(z)  \; ,
\end{align}
which is plotted on the right-hand side of Fig.~\ref{fig-KLM-3}. 

For mesons in category two, Eqs.~(\ref{Mes2-inf1})-(\ref{last-2}), similar angular integrals appear, which can be shown to give vanishing contributions for 
large $J$ by repeating the same steps as above. Hence, by analyzing the angular integrals of the meson BSEs 
we find that transverse gluons have a 
vanishing effect on $\overline{q} q$-bound states for large angular excitation and thus do not alter the symmetry patterns at $J \rightarrow \infty$. 
All mesons with the same $J$ fall into the expected chiral multiplets, Eq.~(\ref{chir-mul}). 

\section{Summary and Conclusions}\label{XIIsum}
In the previous work, Refs.~\cite{Wagenbrunn:2006cs, Wagenbrunn:2007ie}, a very fast effective chiral restoration with increasing  
of the meson spin was established in a chirally symmetric model, that relies on the linear instantaneous
Lorentz-vector confining potential of the Coulomb type. Chiral symmetry in the vacuum is broken
in the standard way via the gap equation. Consequently, the quarks acquire a dynamical momentum-dependent 
mass that decreases at higher momenta. At larger spin of a meson the typical momenta of quarks 
in the meson increase because of the centrifugal repulsion in the system. Hence, the mass of the meson 
is determined by the dynamical quark mass at larger momenta. The larger the typical momentum of quarks in the meson,
the smaller the effect of chiral symmetry breaking in the vacuum on the meson mass and the wave function. This leads to the chiral restoration 
in mesons at large $J$. 

In the present paper we have extended the model and included the effect of transverse gluons. The transverse
gluons influence the dynamics in two ways. First, they enforce chiral symmetry breaking in the vacuum and shift the chiral 
condensate towards a more realistic value, Ref.~\cite{Pak:2011wu}. Still the dynamical mass of quarks decreases very fast with momentum
and vanishes at large momenta. Second, they contribute also to the kernel of the Bethe--Salpeter equations 
for mesons. We have shown that this additional contribution of transverse gluons in meson masses dies out very fast
with increasing $J$. Hence at $J \rightarrow \infty$ the effect of transverse gluons on the meson masses vanishes and 
chiral symmetry is restored in the same way as without transverse gluons. The physical reason is 
that transverse gluons influence the interaction between quarks only at short distances, while the centrifugal repulsion
at large $J$ cuts off the meson wave function at short distances. 
  
\acknowledgements Discussions with R.~F.~Wagenbrunn are greatly acknowledged. 
This work was supported by the Austrian Science Fund (FWF)
through the grant P21970-N16.

\begin{appendix}
\section{Spherical Harmonics and Irreducible Tensor Relations}
\label{tensor-rel-chap}
The irreducible tensor product $\{M_{J_1} \otimes N_{J_2}\}_{J M}$ of two
irreducible tensors $M_{J_1}$ and $N_{J_2}$ is
defined as \cite{Varsh}
\begin{align}
 \{M_{J_1} \otimes N_{J_2}\}_{J M} = \sum_{m_1, m_2} C^{JM}_{J_1 m_1 J_2 m_2} M_{J_1 m_1} N_{J_2 m_2} \; ,
\end{align}
with $m_1 =-J_1, \cdots, J_1, m_2 = -J_2, \cdots, J_2$ and $C^{JM}_{J_1 m_1 J_2 m_2}$ the Clebsch-Gordan coefficients. For instance,
$\{Y_1(\hat{\bp}) \otimes \bg\}_{0} = - \frac{1}{\sqrt{4\pi}} \bg \cdot \hat{\bp}$. The scalar product of 
two irreducible tensors ins defined as
\begin{align}
 M_J \cdot N_J = \sum_{M} (-1)^M \, M_{J M} \, N_{J -M} \; .
\end{align}
In addition to the irreducible tensor relations listed in Ref.~\cite{Wagenbrunn:2007ie} we here use
the following relations which appear for a transverse gluon interaction  of the form (\ref{Ham-trans}) and can be
derived using appropriate formulas in Ref.~\cite{Varsh}: ($\bk = \bp -\bq, \hat{\bk} = \bk/|\bk|$)
\begin{align}
\label{tensor-rel1}
 \frac{4\pi}{2J+1} \{Y_{J_1}(\hat{\bp}) \otimes \hat{\bk} \}_{J} \cdot Y_{J}(\hat{\bq}) =
 \begin{cases} \sqrt{\frac{J}{2J+1}}   \frac{1}{|\bk|}\ \left( |\bp|
P_{J}(\hat{\bp} \cdot \hat{\bq}) - |\bq| P_{J-1}(\hat{\bp} \cdot \hat{\bq}) \right) , & J_1=J-1\\-
 \sqrt{\frac{J+1}{2J+1}} \frac{1}{|\bk|}\left( |\bp| P_{J}(\hat{\bp} \cdot \hat{\bq}) - |\bq| P_{J+1}(\hat{\bp} \cdot \hat{\bq}) \right) , &J_1=J+1
\end{cases}
\end{align}

\begin{align}
 \frac{4\pi}{2J+1} &\{Y_{J_1}(\hat{\bp}) \otimes \hat{\bk} \}_{J} \cdot \{Y_{J_2}(\hat{\bq}) \otimes \hat{\bq} \}_{J} \nonumber \\
& = \begin{cases} \frac{J}{2J+1}  \frac{1}{|\bk|} \left( |\bp|
P_{J}(\hat{\bp} \cdot \hat{\bq}) - |\bq| P_{J-1}(\hat{\bp} \cdot \hat{\bq}) \right) , & J_1=J_2=J-1\\ 
- \frac{\sqrt{J(J+1)}}{2J+1} \frac{1}{|\bk|} \left( |\bp|
P_{J}(\hat{\bp} \cdot \hat{\bq}) - |\bq| P_{J-1}(\hat{\bp} \cdot \hat{\bq}) \right) , &J_1=J-1, J_2=J+1\\
- \frac{\sqrt{J(J+1)}}{2J+1}  \frac{1}{|\bk|} \left( |\bp|  P_{J}(\hat{\bp} \cdot \hat{\bq}) -|\bq| P_{J+1}(\hat{\bp} \cdot \hat{\bq}) \right) ,
&J_1=J+1, J_2=J-1\\
 \frac{J+1}{2J+1} \frac{1}{|\bk|} \left( |\bp| P_{J}(\hat{\bp} \cdot \hat{\bq}) - |\bq| P_{J+1}(\hat{\bp} \cdot \hat{\bq}) \right) ,
&J_1=J_2=J+1\\
\end{cases}
\end{align}

\begin{align}
 &\frac{4\pi}{2J+1} \{Y_{J_1}(\hat{\bp}) \otimes \hat{\bk} \}_{J} \cdot \{Y_{J_2}(\hat{\bq}) \otimes \hat{\bk} \}_{J} \nonumber \\
 & = \begin{cases} \frac{J}{2J+1} \frac{1}{|\bk|^2}  \Big(
(|\bp|^2+|\bq|^2) P_{J-1}(\hat{\bp} \cdot \hat{\bq}) - |\bp||\bq|
(\frac{2J+1}{J} (\hat{\bp} \cdot \hat{\bq})  P_{J-1}(\hat{\bp}
\cdot \hat{\bq}) 
 - \frac{1}{J} P_{J}(\hat{\bp} \cdot \hat{\bq})) \Big) , & J_1=J_2=J-1\\ 
- \frac{\sqrt{J (J+1)}}{2J+1}
\frac{1}{|\bk|^2} 
\Big(|\bp|^2 P_{J+1}(\hat{\bp} \cdot \hat{\bq}) - 2 |\bp||\bq| P_{J}(\hat{\bp} \cdot \hat{\bq}) + |\bq|^2 P_{J-1}(\hat{\bp} \cdot \hat{\bq})\Big) , &J_1=J-1, J_2=J+1\\
- \frac{\sqrt{J(J+1)}}{2J+1}
\frac{1}{|\bk|^2}\Big( |\bp|^2 P_{J-1}(\hat{\bp} \cdot \hat{\bq}) - 2 |\bp||\bq| P_J(\hat{\bp} \cdot \hat{\bq}) +
 |\bq|^2 P_{J+1}(\hat{\bp} \cdot \hat{\bq})  \Big) ,
&J_1=J+1, J_2=J-1\\
\frac{J+1}{2J+1} \frac{1}{|\bk|^2}  \Big(
(|\bp|^2+|\bq|^2) P_{J+1}(\hat{\bp} \cdot \hat{\bq}) - |\bp||\bq|
(\frac{2J+1}{J+1} (\hat{\bp} \cdot \hat{\bq})  P_{J+1}(\hat{\bp}
\cdot \hat{\bq}) + \frac{1}{1+J} P_{J}(\hat{\bp} \cdot \hat{\bq})) \Big) ,
&J_1=J_2=J+1\\
\frac{1}{2J+1} \frac{|\bp||\bq|}{|\bk|^2}
(P_{J+1}(\hat{\bp} \cdot \hat{\bq}) - P_{J-1}(\hat{\bp} \cdot \hat{\bq})) , & J_1=J_2 = J
\end{cases}
\end{align}

\begin{align}
 & \frac{4\pi}{2J+1} i \sum_{j,k,l=1}^{3} \epsilon_{j k l} \{ Y_{J_1}(\hat{\bp}) \otimes \hat{\boldsymbol{e}}_j \}_{J}  \cdot  
\{Y_{J_2}(\hat{\bq}) \otimes \hat{\boldsymbol{e}}_k \}_{J} \frac{k_l}{|\bk|}  \nonumber \\
 &= \begin{cases} \sqrt{\frac{J+1}{2J+1}} \frac{1}{|\bk|} \left( |\bp| P_{J}(\hat{\bp} \cdot \hat{\bq}) - |\bq| P_{J-1}(\hat{\bp} \cdot \hat{\bq}) \right) , & J_1=J-1, J_2=J\\ 
- \sqrt{\frac{J+1}{2J+1}} \frac{1}{|\bk|}\left(|\bp| P_{J-1}(\hat{\bp} \cdot \hat{\bq}) - |\bq| P_{J}(\hat{\bp} \cdot \hat{\bq}) \right) ,
&J_1=J, J_2=J-1\\
- \sqrt{\frac{J}{2J+1}} \frac{1}{|\bk|}\left( |\bp| P_{J+1}(\hat{\bp} \cdot \hat{\bq}) -
|\bq|  P_J(\hat{\bp} \cdot \hat{\bq}) \right) ,
&J_1=J, J_2=J+1\\
\sqrt{\frac{J}{2J+1}} \frac{1}{|\bk|}
\left( |\bp| P_{J}(\hat{\bp} \cdot \hat{\bq}) - |\bq| P_{J+1}(\hat{\bp} \cdot  \hat{\bq}) \right) , & J_1=J+1, J_2 = J
\end{cases}
\end{align}

\begin{align}
\label{tensor-rel4}
 & \frac{4\pi}{2J+1} i \sum_{j,k,l=1}^{3} \epsilon_{j k l}  \{ Y_{J_1}(\hat{\bp}) \otimes \hat{\boldsymbol{e}}_j \}_{J}  \cdot   
 \{Y_{J_2}(\hat{\bq}) \otimes \hat{\bk} \}_{J} \frac{q_k}{|\bq|} \frac{k_l}{|\bk|} \nonumber \\
 &= \begin{cases} \sqrt{\frac{J+1}{2J+1}} \frac{1}{J} \frac{|\bp|^2}{|\bk|^2} \, \left( P_{J+1}(\hat{\bp} \cdot \hat{\bq}) -  (\hat{\bp} \cdot \hat{\bq}) P_{J}(\hat{\bp} \cdot \hat{\bq}) \right) , & J_1=J-1, J_2=J\\ 
  \sqrt{\frac{J+1}{2J+1}} \frac{|\bp|}{|\bk|^2}\Big( |\bp| \left( P_J(\hat{\bp} \cdot \hat{\bq}) - (\hat{\bp} \cdot \hat{\bq}) P_{J-1}(\hat{\bp} \cdot \hat{\bq}) \right)
+ |\bq| \left( (\hat{\bp} \cdot \hat{\bq}) P_{J}(\hat{\bp} \cdot \hat{\bq}) -  P_{J+1}(\hat{\bp} \cdot \hat{\bq}) \right) \Big) ,
&J_1=J, J_2=J-1\\
 \sqrt{\frac{J}{2J+1}} \! \frac{|\bp|}{|\bk|^2} \Big( |\bp|  \left( P_{J}(\hat{\bp} \cdot \hat{\bq}) -
(\hat{\bp} \cdot \hat{\bq})  P_{J+1}(\hat{\bp} \cdot \hat{\bq}) \right) + |\bq| 
\left(  (\hat{\bp} \cdot \hat{\bq})P_{J}(\hat{\bp} \cdot \hat{\bq}) - P_{J-1}(\hat{\bp} \cdot \hat{\bq})  \right) \Big) ,
&J_1=J, J_2=J+1\\
 \sqrt{\frac{J}{2J+1}} \frac{1}{J} \frac{|\bp|^2}{|\bk|^2}
\left( P_{J+1}(\hat{\bp} \cdot  \hat{\bq}) - (\hat{\bp} \cdot  \hat{\bq})  P_{J}(\hat{\bp} \cdot  \hat{\bq}) \right) . & J_1=J+1, J_2 = J
\end{cases}
\end{align}

\section{Coupled Integral Equations}
\label{Appendix-coupled} 
The independent tensor components for an instantaneous interaction are derived in Ref.~\cite{Wagenbrunn:2007ie}. We list the them for each category. 
Mesons of category one are described by
\begin{align}
\label{Cat1}
 \chi_{JM}(\mu, \bp) = \gamma_5 Y_{JM}(\hat{\bp}) \chi_1(\bp) + 
\mu \gamma_0 \gamma_5 Y_{JM}(\hat{\bp}) \chi_2(\bp) &+ 
\mu \gamma_0 \gamma_5 \{ Y_{J+1}(\hat{\bp}) \otimes \bg \}_{J M} \chi_3(\bp)   \nonumber \\
&+ \mu \gamma_0 \gamma_5 \{ Y_{J-1}(\hat{\bp}) \otimes \bg \}_{J M} \chi_4(\bp) \; , 
\end{align}
with $Y_{JM}(\hat{\bp})$ denoting the spherical harmonics. 
The vertex function for mesons of category two is decomposed as
\begin{align}
\label{Cat2}
  \chi_{JM}(\mu, \bp) =  &Y_{JM}(\hat{\bp}) \chi_1(\bp) + 
 \{Y_{J+1}(\hat{\bp}) \otimes \bg \}_{JM} \chi_2(\bp) + 
\{ Y_{J-1}(\hat{\bp}) \otimes \bg \}_{J M} \chi_3(\bp)   \nonumber \\
&+ \mu \gamma_5 \{ Y_{J}(\hat{\bp}) \otimes \bg \}_{J M} \chi_4(\bp) + \mu \gamma_0 \{Y_{J+1}(\hat{\bp}) \otimes \bg \}_{JM} \chi_5(\bp)  
+ \mu \gamma_0 \{Y_{J-1}(\hat{\bp}) \otimes \bg \}_{JM} \chi_6(\bp) \; ,
\end{align}
and for mesons in category three we have
\begin{align}
\label{Cat3}
  \chi_{JM}(\mu, \bp) = \mu \{ Y_{J}(\hat{\bp}) \otimes \bg \}_{J M} \chi_1(\bp) + 
\gamma_5 \{Y_{J+1}(\hat{\bp}) \otimes \bg \}_{JM} \chi_2(\bp) &+ 
\gamma_5 \{ Y_{J-1}(\hat{\bp}) \otimes \bg \}_{J M} \chi_3(\bp)  \nonumber \\
&+ \gamma_0 \{ Y_{J}(\hat{\bp}) \otimes \bg \}_{J M} \chi_4(\bp) \; . 
\end{align}
The coupled integral equations are derived using standard techniques by plugging the independent tensor components into the left- and right-hand 
side of Eq.~(\ref{BSE}). By applying appropriate traces of Dirac matrices and the 
irreducible tensor relations listed in Eqs.~(\ref{tensor-rel1})-(\ref{tensor-rel4}) and \cite{Wagenbrunn:2007ie} we end up 
with a system of coupled integral equations for each category. 

In the following subsections it is shown that the same linear combinations of vertex functions
$\chi_i$ appear for color-Coulomb and transverse gluon interactions for all meson categories. 
By choosing appropriate (infrared finite) linear combinations of these vertex functions, the number of 
coupled equations is reduced. 

\subsection{Mesons of Category One}
For mesons in category $1$ we end up with
a coupled system of four integral equations, given as
\begin{subequations}
\begin{align}
 \chi_1(\bp) = & \, \frac{1}{2} \int \dbar^3 q \frac{V_{\textsc{C}}(\bk)+2 \, D(\bk)}{\omega(\bq) \left(\omega^2(\bq)-\frac{\mu^2}{4}\right)} 
P_{J}(\hat{\bp} \cdot \hat{\bq}) 
\nonumber \\ &\times 
\Bigg[ \omega^2(\bq) \chi_1(\bq)  + \frac{1}{2} \mu^2 A(\bq) \chi_2(\bq)  -  \frac{1}{2} \mu^2 B(\bq) \left( 
\sqrt{\frac{J+1}{2J+1}}  \chi_3 (\bq) -
\sqrt{\frac{J}{2J+1}}  \chi_4 (\bq) \right)
 \Bigg] \; , \\
 \chi_2(\bp) = & \, \frac{1}{2} \int \dbar^3 q \frac{V_{\textsc{C}}(\bk)-2 \, D(\bk)}{\omega^2(\bq)-\frac{\mu^2}{4}} P_{J}(\hat{\bp} \cdot \hat{\bq}) 
\frac{A(\bq)}{\omega(\bq)} \Bigg[ \frac{1}{2} \chi_1(\bq) +
A(\bq) \chi_2(\bq) - B(\bq) \left( \sqrt{\frac{J+1}{2J+1}} \chi_3
(\bq) - \sqrt{\frac{J}{2J+1}} \chi_4 (\bq)  \right) \Bigg] \; , \\
\chi_3(\bp) = & \, - \frac{1}{2} \sqrt{\frac{J+1}{2J+1}} \int \dbar^3 q \frac{V_{\textsc{C}}(\bk) P_{J+1}(\hat{\bp} \cdot \hat{\bq}) - 2 \,
D(\bk) \left( |\bp|
P_{J}(\hat{\bp} \cdot \hat{\bq}) - |\bq| P_{J+1}(\hat{\bp} \cdot \hat{\bq}) \right) \frac{(\hat{\bk} \cdot \hat{\bq})}{|\bk|} }{\omega^2(\bq)-\frac{\mu^2}{4}}  \nonumber \\
&\times \frac{B(\bq)}{\omega(\bq)} \Bigg[\frac{1}{2} \chi_1(\bq) +
A(\bq) \chi_2(\bq) - B(\bq) \left( \sqrt{\frac{J+1}{2J+1}} \chi_3
(\bq) -
 \sqrt{\frac{J}{2J+1}} \chi_4 (\bq) \right)
\Bigg] \; ,  \\
\chi_4(\bp) = & \, \frac{1}{2} \sqrt{\frac{J}{2J+1}} \int \dbar^3 q \frac{V_{\textsc{C}}(\bk)P_{J-1}(\hat{\bp} \cdot \hat{\bq}) - 2\, D(\bk) \left( |\bp|
P_{J}(\hat{\bp} \cdot \hat{\bq}) - |\bq| P_{J-1}(\hat{\bp} \cdot \hat{\bq}) \right) \frac{(\hat{\bk} \cdot \hat{\bq})}{|\bk|}}{\omega^2(\bq)-\frac{\mu^2}{4}}
 \nonumber \\
&\times \frac{B(\bq)}{\omega(\bq)} \Bigg[\frac{1}{2} \chi_1(\bq) +
A(\bq) \chi_2(\bq) - B(\bq) \left( \sqrt{\frac{J+1}{2J+1}} \chi_3
(\bq) -
 \sqrt{\frac{J}{2J+1}} \chi_4 (\bq) \right)
\Bigg] \; . 
\end{align}
\end{subequations}
Setting the transverse gluon interaction to zero, $D(\bk) = 0$, the equations coincide with the result given in Ref.~\cite{Wagenbrunn:2007ie}.
There are only two different linear combinations of the vertex components $\chi_i(\bp)$.
Introducing the two infrared-finite functions (see Ref.~\cite{Wagenbrunn:2007ie})
\begin{align}
 h(\bp) &= \frac{\chi_1(\bp)}{\omega(\bp)}  \; , \\
g(\bp) &= \frac{1}{\omega^2(\bp)-\frac{\mu^2}{4}} \,
\left[\frac{h(\bp)}{\omega(\bp)} + 2 A(\bp) \chi_2(\bp) - 2 B(\bq)
\left(\sqrt{\frac{J+1}{2J+1}} \chi_3(\bp) - \sqrt{\frac{J}{2J+1}}
\chi_4(\bp) \right) \right] \; , 
\end{align}
one ends up with a system of two coupled integral equations
\begin{subequations}
\begin{align}
\label{rev-1}
 \omega(\bp) h(\bp) &= \frac{1}{2} \int \dbar^3 q \, \left( V_{\textsc{C}}(\bk)+2 \, D(\bk) \right) P_{J}(\hat{\bp} \cdot \hat{\bq})
\left[ h(\bq) + \frac{\mu^2}{4 \omega(\bq)} g(\bq) \right] \; , \\
\label{rev-2}
\left[ \omega(\bp) - \frac{\mu^2}{4 \omega(\bp)} \right] g(\bp) &= h(\bp) \nonumber \\
+ \frac{1}{2} \int \dbar^3 q V_{\textsc{C}}(\bk) &\left\{
\frac{A(\bp) A(\bq) P_{J}(\hat{\bp} \cdot \hat{\bq}) + B(\bp)
B(\bq)
\left(\frac{J+1}{2J+1} P_{J+1}(\hat{\bp} \cdot \hat{\bq})  + \frac{J}{2J+1} P_{J-1}(\hat{\bp} \cdot \hat{\bq}) \right)}{\omega(\bp) \omega(\bq)} \right\} g(\bq)  \nonumber \\
- \frac{1}{2} \int \dbar^3 q \, 2 \, D(\bk) &\left\{
\frac{A(\bp)A(\bq) P_{J}(\hat{\bp} \cdot \hat{\bq}) + B(\bp)B(\bq)
\left(\frac{J+1}{2J+1} F_1  + \frac{J}{2J+1}
G_1 \right)}{\omega(\bp) \omega(\bq)} \right\} g(\bq) \; , 
\end{align}
\end{subequations}
where we have used the definitions (\ref{Def-F1}), (\ref{Def-G1}).
For the special case of pions
($J=0$) the equations (\ref{Mes1-1}), (\ref{Mes1-2}) simplify as
\begin{subequations}
\begin{align}
 \omega(\bp) h(\bp) &= \frac{1}{2} \int \dbar^3 q \, \left( V_{\textsc{C}}(\bk)+2 \, D(\bk) \right)
\left[ h(\bq) + \frac{\mu^2}{4 \omega(\bq)} g(\bq) \right] \; , \\
 \left[ \omega(\bp) - \frac{\mu^2}{4 \omega(\bp)} \right] g(\bp) &= h(\bp)  \nonumber \\
+ \frac{1}{2} \int \dbar^3 q V_{\textsc{C}}(\bk) &\left\{
\frac{A(\bp) A(\bq) + B(\bp) B(\bq)
(\hat{\bp} \cdot \hat{\bq})}{\omega(\bp) \omega(\bq)} \right\} g(\bq)  \nonumber \\
- \frac{1}{2} \int \dbar^3 q \, 2 \, D(\bk) & \left\{
\frac{A(\bp)A(\bq)  + B(\bp)B(\bq) (\hat{\bk} \cdot \hat{\bq})
(\hat{\bk} \cdot \hat{\bp})}{\omega(\bp) \omega(\bq)} \right\}
g(\bq) \; ,
\end{align}
\end{subequations}
and agree with the results obtained in
Ref.~\cite{Alkofer:1988tc}. The infrared finite BSEs for mesons of this type are collected in Eqs.~(\ref{Mes1-inf1}), (\ref{Mes1-inf2}).

\subsection{Mesons of Category Two}
For mesons of category two we use the vertex functions (\ref{Cat2}) and have a system of six coupled
integral equations:
\begin{subequations}
\begin{align}
\label{Mes2-A1}
\chi_1(\bp) &= \frac{1}{2} \int \dbar^3 q
\frac{V_{\textsc{C}}(\bk)+ 2
D(\bk)}{\omega^2(\bq)-\frac{\mu^2}{4}} P_{J}(\hat{\bp} \cdot
\hat{\bq}) \frac{B(\bq)}{\omega(\bq)}
 \nonumber \\
& \times \Bigg\{ B(\bq) \chi_1(\bq)  + A(\bq) \left( \sqrt{\frac{J+1}{2J+1}}\chi_2(\bq)  - \sqrt{\frac{J}{2J+1}}  \chi_3(\bq)\right) 
+ \frac{\mu^2}{2} \left(  \sqrt{\frac{J+1}{2J+1}} \chi_5(\bq) -
 \sqrt{\frac{J}{2J+1}} \chi_6(\bq)
\right) \Bigg\}  \; ,  \\
  \chi_2(\bp) &= \frac{1}{2} \int \dbar^3 q \frac{V_{\textsc{C}}(\bk)}{\omega^2(\bq)-\frac{\mu^2}{4}} P_{J+1}(\hat{\bp} \cdot \hat{\bq}) 
\Bigg\{ \omega(\bq) \sqrt{\frac{J}{2J+1}} \left(\sqrt{\frac{J}{2J+1}}\chi_2(\bq) +  \sqrt{\frac{J+1}{2J+1}} \chi_3 (\bq) \right)   \nonumber \\
& + \omega(\bq) \frac{A(\bq)}{\omega(\bq)} \sqrt{\frac{J+1}{2J+1}}
\left[ \frac{B(\bq)}{\omega(\bq)} \chi_1(\bq) +
\frac{A(\bq)}{\omega(\bq)} \left(\sqrt{\frac{J+1}{2J+1}}
\chi_2(\bq)
- \sqrt{\frac{J}{2J+1}} \chi_3(\bq)  \right) \right]  \nonumber \\
& + \frac{\mu^2}{2} \sqrt{\frac{J}{2J+1}}
\left[\frac{B(\bq)}{\omega(\bq)} \chi_4(\bq) +
\frac{A(\bq)}{\omega(\bq)} \left(\sqrt{\frac{J}{2J+1}} \chi_5(\bq)
+ \sqrt{\frac{J+1}{2J+1}} \chi_6(\bq)  \right) \right]  \nonumber \\
& + \frac{\mu^2}{2} \frac{A(\bq)}{\omega(\bq)}
\sqrt{\frac{J+1}{2J+1}} \left(\sqrt{\frac{J+1}{2J+1}} \chi_5(\bq)
- \sqrt{\frac{J}{2J+1}} \chi_6(\bq)  \right) \Bigg\}  \nonumber  \displaybreak  \\
& - \frac{1}{2} \int \dbar^3 q \frac{2 \, D(\bk)}{ \omega^2(\bq)-\frac{\mu^2}{4}} 
 \Bigg\{ \omega(\bq) \sqrt{\frac{J}{2J+1}} F_2 \left(\sqrt{\frac{J}{2J+1}} \chi_2(\bq) + \sqrt{\frac{J+1}{2J+1}} \chi_3(\bq) \right)  \nonumber \\
& - \omega(\bq) \frac{A(\bq)}{\omega(\bq)} \sqrt{\frac{J+1}{2J+1}}
F_1 \left[\frac{B(\bq)}{\omega(\bq)} \chi_1(\bq) +
\frac{A(\bq)}{\omega(\bq)} \left(\sqrt{\frac{J+1}{2J+1}}
\chi_2(\bq)
- \sqrt{\frac{J}{2J+1}} \chi_3(\bq)  \right) \right]  \nonumber \\
& + \frac{\mu^2}{2} \sqrt{\frac{J}{2J+1}} F_2 \left[
\frac{B(\bq)}{\omega(\bq)} \chi_4(\bq) +
\frac{A(\bq)}{\omega(\bq)} \left(\sqrt{\frac{J}{2J+1}} \chi_5(\bq)
+  \sqrt{\frac{J+1}{2J+1}} \chi_6(\bq)  \right) \right]  \nonumber \\
& - \frac{\mu^2}{2} \frac{A(\bq)}{\omega(\bq)}
\sqrt{\frac{J+1}{2J+1}} F_1
\left(\sqrt{\frac{J+1}{2J+1}} \chi_5(\bq) - \sqrt{\frac{J}{2J+1}}
\chi_6(\bq)  \right) \Bigg\} \; ,  \\
 \chi_3(\bp) &= \frac{1}{2} \int \dbar^3 q \frac{V_{\textsc{C}}(\bk)}{\omega^2(\bq)-\frac{\mu^2}{4}} P_{J-1}(\hat{\bp} \cdot \hat{\bq}) 
 \Bigg\{ \omega(\bq) \sqrt{\frac{J+1}{2J+1}} \left(\sqrt{\frac{J}{2J+1}}\chi_2(\bq) +  \sqrt{\frac{J+1}{2J+1}} \chi_3 (\bq) \right)  \nonumber \\
& - \omega(\bq) \frac{A(\bq)}{\omega(\bq)} \sqrt{\frac{J}{2J+1}}
\left[ \frac{B(\bq)}{\omega(\bq)} \chi_1(\bq) +
\frac{A(\bq)}{\omega(\bq)} \left(\sqrt{\frac{J+1}{2J+1}}
\chi_2(\bq)
- \sqrt{\frac{J}{2J+1}} \chi_3(\bq)  \right) \right]  \nonumber \\
& + \frac{\mu^2}{2} \sqrt{\frac{J+1}{2J+1}}
\left[\frac{B(\bq)}{\omega(\bq)} \chi_4(\bq) +
\frac{A(\bq)}{\omega(\bq)} \left(\sqrt{\frac{J}{2J+1}} \chi_5(\bq)
+ \sqrt{\frac{J+1}{2J+1}} \chi_6(\bq)  \right) \right]  \nonumber \\
& - \frac{\mu^2}{2} \frac{A(\bq)}{\omega(\bq)}
\sqrt{\frac{J}{2J+1}} \left(\sqrt{\frac{J+1}{2J+1}} \chi_5(\bq)
- \sqrt{\frac{J}{2J+1}} \chi_6(\bq)  \right) \Bigg\}  \nonumber \\
& - \frac{1}{2} \int \dbar^3 q \frac{2 \, D(\bk)}{\omega^2(\bq)-\frac{\mu^2}{4}} 
 \Bigg\{ \omega(\bq)
 \sqrt{\frac{J(J+1)}{2J+1}} G_2  \left(\sqrt{\frac{J}{2J+1}}\chi_2(\bq) +  \sqrt{\frac{J+1}{2J+1}} \chi_3 (\bq) \right)
 \nonumber \\
& + \omega(\bq) \frac{A(\bq)}{\omega(\bq)} \sqrt{\frac{J}{2J+1}}
G_1 \left[\frac{B(\bq)}{\omega(\bq)} \chi_1(\bq) +
\frac{A(\bq)}{\omega(\bq)} \left(\sqrt{\frac{J+1}{2J+1}}
\chi_2(\bq)
- \sqrt{\frac{J}{2J+1}} \chi_3(\bq)  \right) \right]  \nonumber \\
& + \frac{\mu^2}{2} \sqrt{\frac{J+1}{2J+1}} G_2
\left[ \frac{B(\bq)}{\omega(\bq)} \chi_4(\bq) +
\frac{A(\bq)}{\omega(\bq)} \left(\sqrt{\frac{J}{2J+1}} \chi_5(\bq)
+ \sqrt{\frac{J+1}{2J+1}} \chi_6(\bq)  \right) \right]  \nonumber
\\ & + \frac{\mu^2}{2} \frac{A(\bq)}{\omega(\bq)}
\sqrt{\frac{J}{2J+1}} G_1 \left(\sqrt{\frac{J+1}{2J+1}}
 \chi_5(\bq) - \sqrt{\frac{J}{2J+1}}
 \chi_6(\bq) \right) \Bigg\} \; ,\\
 \chi_4(\bp) &= \frac{1}{2} \int \dbar^3 q \frac{V_{\textsc{C}}(\bk)}{\omega^2(\bq)-\frac{\mu^2}{4}} P_{J}(\hat{\bp} \cdot \hat{\bq}) 
 \Bigg\{\frac{1}{2} \frac{B(\bq)}{\omega(\bq)} \left(\sqrt{\frac{J}{2J+1}} \chi_2(\bq) + \sqrt{\frac{J+1}{2J+1}} \chi_3(\bq)\right)  \nonumber \\
& +  \omega(\bq) \frac{B(\bq)}{\omega(\bq)}
\left[\frac{B(\bq)}{\omega(\bq)} \chi_4(\bq) +
\frac{A(\bq)}{\omega(\bq)}
\left( \sqrt{\frac{J}{2J+1}} \chi_5(\bq) + \sqrt{\frac{J+1}{2J+1}} \chi_6(\bq)\right) \right] \Bigg\}  \nonumber \\
& + \frac{1}{2} \int \dbar^3 q \frac{2 \,
D(\bk)}{\omega^2(\bq)-\frac{\mu^2}{4}}
\frac{H}{J} 
 \Bigg\{ \frac{1}{2} \frac{B(\bq)}{\omega(\bq)} \left(\sqrt{\frac{J}{2J+1}} \chi_2(\bq) + \sqrt{\frac{J+1}{2J+1}} \chi_3(\bq)\right)   \nonumber \\
& +  \omega(\bq) \frac{B(\bq)}{\omega(\bq)}
\left[\frac{B(\bq)}{\omega(\bq)} \chi_4(\bq) +
\frac{A(\bq)}{\omega(\bq)} \left( \sqrt{\frac{J}{2J+1}}
\chi_5(\bq) + \sqrt{\frac{J+1}{2J+1}} \chi_6(\bq)\right) \right]
\Bigg\} \; ,    \\
\chi_5(\bp) &= \frac{1}{2} \int \dbar^3 q \frac{V_{\textsc{C}}(\bk)}{\omega^2(\bq)-\frac{\mu^2}{4}} P_{J+1}(\hat{\bp} \cdot \hat{\bq}) 
 \Bigg\{
\frac{1}{2} \frac{A(\bq)}{\omega(\bq)} \sqrt{\frac{J}{2J+1}} \left(\sqrt{\frac{J}{2J+1}} \chi_2(\bq) + \sqrt{\frac{J+1}{2J+1}} \chi_3(\bq) \right)  \nonumber \\
& +  \frac{1}{2} \sqrt{\frac{J+1}{2J+1}} \left[
\frac{B(\bq)}{\omega(\bq)} \chi_1(\bq)  +
\frac{A(\bq)}{\omega(\bq)} \left(\sqrt{\frac{J}{2J+1}} \chi_2(\bq) - \sqrt{\frac{J+1}{2J+1}}  \chi_3(\bq) \right)\right]  \nonumber \\
& +  \frac{A(\bq)}{\omega(\bq)} \omega(\bq) \sqrt{\frac{J}{2J+1}}
\left[ \frac{B(\bq)}{\omega(\bq)} \chi_4(\bq)  +
\frac{A(\bq)}{\omega(\bq)} \left(\sqrt{\frac{J}{2J+1}} \chi_5(\bq)  + \sqrt{\frac{J+1}{2J+1}} \chi_6(\bq) \right)\right]  \nonumber \\
& +  \omega(\bq)  \sqrt{\frac{J+1}{2J+1}}  \left(\sqrt{\frac{J+1}{2J+1}} \chi_5(\bq) - \sqrt{\frac{J}{2J+1}}  \chi_6(\bq)\right) \Bigg\}  \nonumber  \displaybreak \\
& - \frac{1}{2} \int \dbar^3 q \frac{2 \, D(\bk)}{\omega^2(\bq)-\frac{\mu^2}{4}} 
 \Bigg\{ - \frac{1}{2} \frac{A(\bq)}{\omega(\bq)}
\sqrt{\frac{J}{2J+1}} F_2 \left(
\sqrt{\frac{J}{2J+1}} \chi_2(\bq) +  \sqrt{\frac{J+1}{2J+1}} \chi_3(\bq)
\right)  \nonumber \\
& +  \frac{1}{2} \sqrt{\frac{J+1}{2J+1}} F_1 \left[
\frac{B(\bq)}{\omega(\bq)} \chi_1(\bq)  +
\frac{A(\bq)}{\omega(\bq)} \left(\sqrt{\frac{J}{2J+1}} \chi_2(\bq) - \sqrt{\frac{J+1}{2J+1}}  \chi_3(\bq) \right)\right]  \nonumber \\
& -  \frac{A(\bq)}{\omega(\bq)} \omega(\bq) \sqrt{\frac{J}{2J+1}}
F_2 \left[ \frac{B(\bq)}{\omega(\bq)} \chi_4(\bq)  +
\frac{A(\bq)}{\omega(\bq)} \left(\sqrt{\frac{J}{2J+1}} \chi_5(\bq) 
+ \sqrt{\frac{J+1}{2J+1}} \chi_6(\bq) \right)\right]  \nonumber \\
& +  \omega(\bq)  \sqrt{\frac{J+1}{2J+1}} F_1
\left(\sqrt{\frac{J+1}{2J+1}} \chi_5(\bq)  - 
\sqrt{\frac{J}{2J+1}} \chi_6(\bq)\right) \Bigg\}  \; ,  \\
\label{Mes2-A6}
\chi_6(\bp) &= \frac{1}{2} \int \dbar^3 q \frac{V_{\textsc{C}}(\bk)}{\omega^2(\bq)-\frac{\mu^2}{4}} P_{J-1}(\hat{\bp} \cdot \hat{\bq}) 
 \Bigg\{
\frac{1}{2} \frac{A(\bq)}{\omega(\bq)} \sqrt{\frac{J+1}{2J+1}} \left(\sqrt{\frac{J}{2J+1}} \chi_2(\bq)  + \sqrt{\frac{J+1}{2J+1}} \chi_3(\bq) \right)  \nonumber \\
& -  \frac{1}{2} \sqrt{\frac{J}{2J+1}} \left[
\frac{B(\bq)}{\omega(\bq)} \chi_1(\bq)  +
\frac{A(\bq)}{\omega(\bq)} \left(\sqrt{\frac{J}{2J+1}} \chi_2(\bq) - \sqrt{\frac{J}{2J+1}} \chi_3(\bq)  \right)\right]  \nonumber \\
& +  \frac{A(\bq)}{\omega(\bq)} \omega(\bq) \sqrt{\frac{J+1}{2J+1}}
\left[ \frac{B(\bq)}{\omega(\bq)} \chi_4(\bq)  +
\frac{A(\bq)}{\omega(\bq)} \left(\sqrt{\frac{J}{2J+1}} \chi_5(\bq)  +  \sqrt{\frac{J+1}{2J+1}} \chi_6(\bq) \right)\right]  \nonumber \\
& -  \omega(\bq)  \sqrt{\frac{J}{2J+1}}  \left(\sqrt{\frac{J+1}{2J+1}} \chi_5(\bq) - \sqrt{\frac{J}{2J+1}} \chi_6(\bq) \right) \Bigg\}  \nonumber \\
& + \frac{1}{2} \int \dbar^3 q \frac{2 \, D(\bk)}{\omega^2(\bq)-\frac{\mu^2}{4}} 
 \Bigg\{
 \frac{1}{2} \frac{A(\bq)}{\omega(\bq)} \sqrt{\frac{J+1}{2J+1}} G_2 \left( \sqrt{\frac{J}{2J+1}} \chi_2(\bq) + \sqrt{\frac{J+1}{2J+1}}  \chi_3(\bq) \right)  \nonumber \\
& +  \frac{1}{2} \sqrt{\frac{J}{2J+1}} G_1
\left[\frac{B(\bq)}{\omega(\bq)} \chi_1(\bq) +
\frac{A(\bq)}{\omega(\bq)} \left( \sqrt{\frac{J+1}{2J+1}} \chi_2(\bq) - \sqrt{\frac{J}{2J+1}} \chi_3(\bq) \right)\right]  \nonumber \\
& +  \frac{A(\bq)}{\omega(\bq)} \omega(\bq) G_2  \left[
\frac{B(\bq)}{\omega(\bq)} \chi_4(\bq)
  + \frac{A(\bq)}{\omega(\bq)} \left(
\sqrt{\frac{J}{2J+1}} \chi_5(\bq)
+ \sqrt{\frac{J+1}{2J+1}} \chi_6(\bq) \right)\right]  \nonumber \\
& +  \omega(\bq)  \sqrt{\frac{J}{2J+1}} G_1
\left(\sqrt{\frac{J+1}{2J+1}} \chi_5(\bq)  - 
\sqrt{\frac{J}{2J+1}} \chi_6(\bq) \right) \Bigg\} \; ,
\end{align}
\end{subequations}
with the definitions (\ref{Def-F1}), (\ref{Def-F2}), (\ref{Def-G2}), (\ref{H}). 
We use the following functions 
\begin{align}
h_1(\bp) & = \frac{\sqrt{\frac{J}{2J+1}} \chi_2(\bp) +
\sqrt{\frac{J+1}{2J+1}}\chi_3(\bp)}{\omega(\bp)} \; , \\
g_1(\bp) & = \frac{\omega(\bp) \left[ h_1(\bp) + 2
\frac{B(\bp)}{\omega(\bp)} \chi_4(\bp) + 2
\frac{A(\bp)}{\omega(\bp)} \left( 
\sqrt{\frac{J}{2J+1}} \chi_5(\bp) +
 \sqrt{\frac{J+1}{2J+1}} \chi_6(\bp) \right) \right]}{\omega^2(\bp)-\frac{\mu^2}{4}} \; , \\
h_2(\bp) & = \frac{\frac{B(\bq)}{\omega(\bp)} \chi_1(\bp) +
\frac{A(\bp)}{\omega(\bp)} \left( \sqrt{\frac{J+1}{2J+1}}
\chi_2(\bp) -
\sqrt{\frac{J}{2J+1}}\chi_3(\bp) \right)}{\omega(\bp)} \; , \\
g_2(\bp) & = \frac{\omega(\bp) \left[ h_2(\bp) + 2
 \left( 
\sqrt{\frac{J+1}{2J+1}} \chi_5(\bp) - \sqrt{\frac{J}{2J+1}} \chi_6(\bp) \right)
\right]}{\omega^2(\bp)-\frac{\mu^2}{4}} \; ,
\end{align}
in order to reduce the Eqs.~(\ref{Mes2-A1})-(\ref{Mes2-A6}) to
a system of four coupled integral equations ($P_J = P_{J}(\hat{\bp} \cdot \hat{\bq})$)
\begin{subequations}
\begin{align}
\label{Mes2-1}\omega(\bp) h_1(\bp) =& \, \frac{1}{2} \int \dbar^3 q
\, V_{\textsc{C}}(\bk) \Bigg\{  \left(\frac{J}{2J+1} P_{J+1} + \frac{J+1}{2J+1} P_{J-1}\right)
\left( h_1(\bq) + \frac{\mu^2}{4 \omega(\bq)} g_1(\bq) \right) 
 \nonumber \\ & \quad \quad 
+ \frac{A(\bq)}{\omega(\bq)}
\frac{\sqrt{J(J+1)}}{2J+1} \left(P_{J+1} -
P_{J-1} \right) \left(h_2(\bq) + \frac{\mu^2}{4 \omega(\bq)} g_2(\bq) \right) \Bigg\}  \nonumber  \displaybreak \\
& \quad - \frac{1}{2}  \int \dbar^3 q  \, 2 D(\bk) \Bigg\{ \left(
\frac{J}{2J+1} F_2
+ \frac{J+1}{2J+1} G_2 \right) \left( h_1(\bq) + \frac{\mu^2}{4 \omega(\bq)} g_1(\bq) \right)  \nonumber \\
& \quad \quad + \frac{A(\bq)}{\omega(\bq)}
\frac{\sqrt{J(J+1)}}{2J+1} \left( F_1 -
G_1 \right) \left(h_2(\bq) + \frac{\mu^2}{4
\omega(\bq)} g_2(\bq) \right) \Bigg\} \; , \\
\label{Mes2-2} \left( \omega(\bp) - \frac{\mu^2}{4 \omega(\bp)}
\right)
g_1(\bp) =& \, h_1(\bp)  \nonumber \\
+ \frac{1}{2} \int \dbar^3 q  & \, V_{\textsc{C}}(\bk)
\Bigg\{ \frac{B(\bp) B(\bq) P_J +
A(\bp)A(\bq)
\left( \frac{J}{2J+1} P_{J+1} + \frac{J+1}{2J+1} P_{J-1} \right)}{\omega(\bp) \omega(\bq)} g_1(\bq)   \nonumber \\
& \quad \quad +  \frac{A(\bp)}{\omega(\bp)} \frac{\sqrt{J (J+1)}}{2J+1} \left( P_{J+1} - P_{J-1} \right) g_2(\bq) \Bigg\}  \nonumber  \\
+ \frac{1}{2} \int \!\dbar^3 q & \, 2 D(\bk) \Bigg\{ \frac{B(\bp)
B(\bq) \frac{H}{J}  + A(\bp)A(\bq)
\left( \frac{J}{2J+1} F_2 + \frac{J+1}{2J+1} G_2 \right)}{\omega(\bp) \omega(\bq)} g_1(\bq)   \nonumber \\
& \quad \quad - \frac{A(\bp)}{\omega(\bp)} \frac{\sqrt{J
(J+1)}}{2J+1} \left( F_1 - G_1 \right)
g_2(\bq) \Bigg\} \; ,   \\
\label{Mes2-3}
\omega(\bp) h_2(\bp) & \, = \nonumber \\
= \frac{1}{2} \int \dbar^3 q  V_{\textsc{C}}(\bk)  \Bigg\{ &
\!\! \frac{B(\bp) B(\bq) P_J \!+\!
A(\bp)A(\bq) \left( \!\frac{J+1}{2J+1} P_{J+1} \!+\! \frac{J}{2J+1} P_{J-1}
\right)}{\omega(\bp) \omega(\bq)}
\left( h_2(\bq) \!+\! \frac{\mu^2}{4 \omega(\bq)} \! g_2(\bq) \right)   \nonumber \\
& \quad \quad + \frac{A(\bq)}{\omega(\bq)} \frac{\sqrt{J
(J+1)}}{2J+1} \left( P_{J+1} -
P_{J-1} \right)
\left( h_1(\bq) + \frac{\mu^2}{4 \omega(\bq)} g_1(\bq) \right) \Bigg\}  \nonumber \\
+ \frac{1}{2} \int \!\dbar^3 q  2 D(\bk) \Bigg\{ & \frac{B(\bp)
B(\bq) P_J + A(\bp)A(\bq)
\left( \frac{J+1}{2J+1} F_1 + \frac{J}{2J+1} G_1 \right)}{\omega(\bp) \omega(\bq)} \left( h_2(\bq) + \frac{\mu^2}{4 \omega(\bq)} g_2(\bq) \right)  \nonumber \\
& \quad \quad - \frac{A(\bq)}{\omega(\bq)} \frac{\sqrt{J
(J+1)}}{2J+1} \left(F_2 - G_2 \right)
\left( h_1(\bq) + \frac{\mu^2}{4 \omega(\bq)} g_1(\bq) \right)
\Bigg\} \; , \\
\label{Mes2-4}
\left( \omega(\bp) - \frac{\mu^2}{4 \omega(\bq)} \right) g_2(\bq) =& \, h_2(\bp)  \nonumber \\
+ \frac{1}{2} \int \!\dbar^3 q  \, V_{\textsc{C}}(\bk) \Bigg\{ & \left( \frac{J+1}{2J+1} P_{J+1} + 
\frac{J}{2J+1} P_{J-1} \right) g_2(\bq) 
+ \frac{A(\bq)}{\omega(\bq)} \frac{\sqrt{J (J+1)}}{2J+1} \left(
P_{J+1} - P_{J-1}  \right)
g_1(\bq) \Bigg\} \nonumber \\
-  \frac{1}{2} \int \!\dbar^3 q 2 D(\bk) \Bigg\{ & \left( \frac{J+1}{2J+1} F_1 + \frac{J}{2J+1} G_1 \right) g_2(\bq) +
\frac{A(\bq)}{\omega(\bq)} \frac{\sqrt{J (J+1)}}{2J+1}
\left(F_2 - G_2  \right) g_1(\bq)
\Bigg\} \; .
\end{align}
\end{subequations}
As pointed out in Section \ref{Bethe} for the state $0^{++}$ only the last two equations apply. 

\subsection{Mesons of Category Three}
For mesons of category three we use the set of functions (\ref{Cat3}), apply the standard projection procedures and arrive at a system of four integral equations:
\begin{subequations}
\begin{align}
\chi_1(\bp) = & \, \frac{1}{2} \int \dbar^3 q
\frac{V_{\textsc{C}}(\bk) P_{J}(\hat{\bp} \cdot \hat{\bq}) + 2 D(\bk)\frac{H}{J}}{\omega(\bq)
\left(\omega^2(\bq)-\frac{\mu^2}{4}\right)}  \nonumber \\
& \times 
 \Bigg[  \omega^2(\bq) \chi_1(\bq) + \frac{1}{2}  B(\bq) \left(
\sqrt{\frac{J}{2J+1}}  \chi_2(\bq) + 
\sqrt{\frac{J+1}{2J+1}}  \chi_3(\bq) \right)  + \frac{1}{2}  A(\bq) \chi_4(\bq)
\Bigg] \; ,  \label{chi_1_mes3}  \displaybreak \\
\chi_2(\bp) = & \, \sqrt{\frac{J}{2J+1}}\frac{1}{2} \int \dbar^3 q
\frac{V_{\textsc{C}}(\bk) P_{J+1}(\hat{\bp} \cdot \hat{\bq}) - 2 \, D(\bk) F_2}{\omega(\bq) \left(\omega^2(\bq)-\frac{\mu^2}{4}\right)}
 \frac{B(\bq)}{\omega(\bq)} \nonumber \\
& \times 
 \Bigg[\frac{1}{2} \mu^2 \chi_1(\bq)   + B(\bq) \left(
\sqrt{\frac{J}{2J+1}} \chi_2(\bq) + \sqrt{\frac{J+1}{2J+1}}
 \chi_3(\bq) \right) + A(\bq) \chi_4 (\bq) 
\Bigg] \; , \label{chi_2-mes3} \\
\chi_3(\bp) = & \, \sqrt{\frac{J+1}{2J+1}} \frac{1}{2} \int \dbar^3 q
\frac{V_{\textsc{C}}(\bk) P_{J-1}(\hat{\bp} \cdot
\hat{\bq}) - 2 \, D(\bk)  G_2 }{\omega(\bq)
\left(\omega^2(\bq)-\frac{\mu^2}{4}\right)}  \frac{B(\bq)}{\omega(\bq)}  \nonumber \\
& \times \Bigg[\frac{1}{2} \mu^2 \chi_1(\bq) + B(\bq) \left(
\sqrt{\frac{J}{2J+1}}  \chi_2(\bq) + \sqrt{\frac{J+1}{2J+1}}
  \chi_3(\bq) \right) +  A(\bq) \chi_4 (\bq)
\Bigg] \; , \label{chi_3-mes3} \\
 \chi_4(\bp) = & \, \frac{1}{2} \int \dbar^3 q \frac{V_{\textsc{C}}(\bk) P_{J} (\hat{\bp} \cdot \hat{\bq}) - 2 \, D(\bk) 
 \frac{H}{J}}{\omega(\bq) \left(\omega^2(\bq)-\frac{\mu^2}{4}\right)} 
  \frac{A(\bq)}{\omega(\bq)}  \nonumber \\
& \times  \Bigg[\frac{1}{2} \mu^2 \chi_1(\bq) + B(\bq) \left(
\sqrt{\frac{J}{2J+1}}  \chi_2(\bq) + \sqrt{\frac{J+1}{2J+1}}
 \chi_3 (\bq) \right) +  A(\bq) \chi_4 (\bq)
\Bigg] \; . \label{chi_4-mes3} 
\end{align}
\end{subequations}
The infrared finite functions
\begin{align}
h(\bp) = \frac{\frac{B(\bp)}{\omega(\bp)} \left( \sqrt{\frac{J}{2J+1}}
\chi_2(\bq) +
 \sqrt{\frac{J+1}{2J+1}} \chi_3(\bq) \right) + \frac{A(\bp)}{\omega(\bp)} \chi_4(\bp)}{\omega(\bp)} \, ,
\end{align}
and
\begin{align}
g(\bp) = \frac{\omega(\bp) \left[ h(\bp) + 2 \chi_1(\bp)
\right]}{\omega^2(\bp) - \frac{\mu^2}{4}} \; , 
\end{align}
fulfill the equations
\begin{subequations}
\begin{align}
\label{Mes3-1}
\left( \omega(\bp) - \frac{\mu^2}{4 \omega(\bp)}
\right) g(\bp) = \, & h(\bp) +
 \frac{1}{2} \int \dbar^3 q \, \left( V_{\textsc{C}}(\bk) P_{J}(\hat{\bp} \cdot \hat{\bq}) +
 2  D(\bk) \frac{H}{J} \right)  g(\bq) \, , \\
\omega(\bp) h(\bp)  \, & \nonumber \\
\label{Mes3-2}
 = \frac{1}{2} \int \dbar^3 q V_{\textsc{C}}(\bk) & \Bigg\{ 
\frac{A(\bp) A(\bq)  P_{J} + B(\bp)
B(\bq)
\left(\frac{J}{2J+1} P_{J+1} + \frac{J+1}{2J+1} P_{J-1}\right)}{\omega(\bp) \omega(\bq)}
\left( h(\bq) + \frac{\mu^2}{4 \omega(\bq)} g(\bq) \right) \Bigg\}  \nonumber \\
- \frac{1}{2} \int \dbar^3 q \, 2 \, D(\bk) & \Bigg\{ 
\frac{A(\bp) A(\bq) \frac{H}{J} + B(\bp) B(\bq) \left(
\frac{J}{2J+1} F_2 + \frac{J+1}{2J+1}
G_2 \right)}{\omega(\bp) \omega(\bq)} \left( h(\bq) +
\frac{\mu^2}{4 \omega(\bq)} g(\bq) \right) \Bigg\} \; ,
\end{align}
\end{subequations}
with $F_2, G_2,H$ given in
Eqs.~(\ref{Def-F2}), (\ref{Def-G2}), (\ref{H}). For $J=0$, which corresponds to a $0^{--}$ meson, we
have only the component $\chi_2(\bp)$, see Eq.~(\ref{Cat3}). However, Eq.~(\ref{chi_2-mes3}) comes
with an overall $\sqrt{\frac{J}{2J+1}}$ factor, so such a meson does not occur in a model with instantaneous interaction,
which goes in line with the findings of Ref.~\cite{Wagenbrunn:2007ie}.

\end{appendix}

\bibliographystyle{utphys}
\bibliography{Biblio2}

\end{document}